\documentclass[%
 aip,
 jcp,
 amsmath,amssymb,
 reprint,%
]{revtex4-1}

\usepackage{graphicx} 
\usepackage{booktabs}
\usepackage{dcolumn}
\usepackage[utf8]{inputenc}
\usepackage{amsmath}
\usepackage{amsfonts}
\usepackage{amssymb}
\usepackage{bbm}
\usepackage[dvipsnames]{xcolor}

\usepackage{multirow}
\usepackage{braket}
\usepackage{hyperref} 
\usepackage[capitalise,nameinlink]{cleveref}
\usepackage{bm}

\usepackage{xspace}
\usepackage{arydshln}
\usepackage{tikz}
\usetikzlibrary{tikzmark}
\usetikzlibrary{arrows, arrows.meta}
\usepackage{comment}
\usepackage{newtxtext}
\usepackage{mathtools}

\newcommand*{\wfqfmu}{$\omega$FQF$\mu$\xspace}

\newcommand*{\wfqfmubem}{$\omega$FQF$\mu$-BEM\xspace}

\newcommand*{\imm}{\mathop{}\!\mathrm{i}}

\newcommand*{\tqq}{\mathbf{T}^\mathrm{qq}}
\newcommand*{\tqmu}{\mathbf{T}^\mathrm{q\mu}}
\newcommand*{\tmuq}{\mathbf{T}^\mathrm{\mu q}}
\newcommand*{\tmumu}{\mathbf{T}^\mathrm{\mu\mu}}
\newcommand*{\tqsigma}{\mathbf{T}^\mathrm{q\sigma}}
\newcommand*{\tmusigma}{\mathbf{T}^\mathrm{\mu\sigma}}
\newcommand*{\tsigmaq}{\mathbf{T}^\mathrm{\sigma q}}
\newcommand*{\tsigmamu}{\mathbf{T}^\mathrm{\sigma\mu}}

\newcommand*{\sm}{SM\xspace}

\usepackage{balance}
\usepackage{mathptmx}
\usepackage{etoolbox}
\usepackage{lastpage}
\usepackage{float}
\usepackage{fancyhdr}
\usepackage{fnpos}
\usepackage[english]{babel}
\usepackage{array}
\usepackage{droidsans}
\usepackage{charter}
\usepackage[T1]{fontenc}

\usepackage{hyperref}

\usepackage{epstopdf}
\AppendGraphicsExtensions{.tiff}

\makeatletter
\def\@email#1#2{%
 \endgroup
 \patchcmd{\titleblock@produce}
  {\frontmatter@RRAPformat}
  {\frontmatter@RRAPformat{\produce@RRAP{*#1\href{mailto:#2}{#2}}}\frontmatter@RRAPformat}
  {}{}
}%
\makeatother

\begin{document}

\preprint{AIP/123-QED}

\title[]{Mixed Atomistic-Implicit Quantum/Classical Approach to Molecular Nanoplasmonics}

\author{Pablo Grobas Illobre}
\affiliation{Scuola Normale Superiore,
             Piazza dei Cavalieri 7, 56126 Pisa, Italy.}
\author{Piero Lafiosca}
\affiliation{Scuola Normale Superiore,
             Piazza dei Cavalieri 7, 56126 Pisa, Italy.} 
\author{Luca Bonatti}
\affiliation{Scuola Normale Superiore,
             Piazza dei Cavalieri 7, 56126 Pisa, Italy.}
\author{Tommaso Giovannini}
\email{tommaso.giovannini@uniroma2.it}
\affiliation{Department of Physics, University of Rome Tor Vergata, Via della Ricerca Scientifica 1, 00133, Rome, Italy.}
\author{Chiara Cappelli}
\email{chiara.cappelli@sns.it}
\affiliation{Scuola Normale Superiore,
             Piazza dei Cavalieri 7, 56126 Pisa, Italy.}
             \affiliation{IMT School for Advanced Studies Lucca, Piazza San Francesco 19, Lucca, 55100, Italy}


\begin{abstract}
A multiscale QM/classical approach is presented, that is able to model the optical properties of complex nanostructures composed of a molecular system adsorbed on metal nanoparticles. The latter are described by a combined atomistic-continuum model, where the core is described using the implicit boundary element method (BEM) and the surface retains a fully atomistic picture and is treated employing the frequency-dependent fluctuating charge and fluctuating dipole (\wfqfmu) approach. 
The integrated QM/\wfqfmubem model is numerically compared with state-of-the-art fully atomistic approaches, and the quality of the continuum/core partition is evaluated. The method is then extended to compute Surface-Enhanced Raman Scattering (SERS) within a Time-Dependent Density Functional Theory (TDDFT) framework.
\end{abstract}

\maketitle

\newpage

\section{Introduction}


When metal nanoparticles (NPs) are irradiated with external radiation, coherent oscillations of conduction electrons, also named localized surface plasmons (LSPs), can be excited.\cite{giannini2011plasmonic,maier2007plasmonics,odom2011introduction} Most NP optical properties are related to LSP peculiar properties, which can be tuned by varying the NP's size, shape, and chemical composition.\cite{kelly2003optical} Plasmonic nanostructures can enhance, control, or suppress properties of molecules interacting with light: these features are exploited in molecular nanoplasmonics.\cite{coccia2020hybrid,willets2007localized,zhang2013chemical,jiang2015distinguishing,chiang2016conformational,langer2019present} A deep understanding of the phenomena that occur at the molecular and nanoscale in the presence of light can be achieved by exploiting multiscale hybrid techniques, which use different levels of description for molecules and plasmonic nanosystems.\cite{coccia2020hybrid,lafiosca2023classical} With these methods, a reliable representation of both atomistic details and collective features, such as plasmons, in these complex systems can be achieved.

In principle, the ideal theoretical approach to molecular nanoplasmonics should rely on a quantum mechanical (QM) description of the whole system (molecule + nanostructured substrate). However, due to their unfavorable computational scaling, full {\it{ab initio}} methods are currently limited to small model systems (generally, $<$ 100 atoms), in contrast to the large size of plasmonic substrates, which typically comprise thousands or millions of atoms.\cite{zhao2006pyridine,jensen2008electronic,morton2009understanding,morton2011theoretical,gonzalez2011surface} To address these limitations, hybrid multiscale QM/classical approaches have been developed,\cite{lafiosca2023classical,morton2010discrete,morton2011discrete} where the molecular adsorbate is treated at the QM level, while the substrate's plasmonic response is calculated by using classical electrodynamical methods, substantially reducing the computational cost. The most crude classical description of the plasmonic response of a metal NP is given by continuum implicit approaches, such as the Mie theory,\cite{mie1908beitrage} the Finite Difference Time Domain (FDTD),\cite{taflove2005computational} or the Boundary Element Method (BEM),\cite{de2002retarded,myroshnychenko2008modeling,mennucci2019multiscale,bonatti2020plasmonic}, which completely disregard the NP atomistic nature. To overcome this limitation, fully atomistic approaches have been developed, such as Discrete Interaction Models (DIMs),\cite{morton2010discrete,morton2011discrete,jensen2008electrostatic,jensen2009atomistic,zakomirnyi2019extended,zakomirnyi2020plasmonic} and the Frequency-Dependent Fluctuating Charges and Fluctuating Dipoles (\wfqfmu) approach.\cite{giovannini2019classical,bonatti2020plasmonic,giovannini2020graphene,lafiosca2021going,bonatti2022silico,giovannini2022we,zanotto2023strain,lafiosca2023classical,nicoli2023fully,lafiosca2024real,nicoli2024atomistic} \wfqfmu can correctly reproduce QM
results for metal NPs even below the quantum limit ($<$ 5 nm),\cite{giovannini2022we} capture the plasmonic properties of systems featuring subnanometer junctions,\cite{giovannini2019classical} defects,\cite{bonatti2022silico,zanotto2023strain} treat bimetallic particles,\cite{nicoli2023fully} and colloidal nanostructures.\cite{nicoli2024atomistic} Also, it describes under the same theoretical framework noble metal NPs\cite{giovannini2022we} and graphene-based structures.\cite{giovannini2020graphene}
Both families (implicit and atomistic) have pros and cons. Continuum models feature a favorable computational cost, which scales with the size of the NP surface, facilitating the description of large systems.\cite{corni2001enhanced,corni2001theoretical,corni2002surface,corni2002erratum,corni2006studying,mennucci2019multiscale,corni2021role,corni2024unraveling} However, their implicit, non-atomistic nature fails at capturing the NP plasmonic response in specific configurations characterized by atomistic defects, sub-nanometer junctions, and sharp interfaces,\cite{corni2021role,trugler2012} which are associated with huge enhancements of the electric field (the so-called hot-spots). Conversely, while atomistic approaches appropriately capture these features, they become computationally less efficient as the size of the system increases, because their computational cost scales with the number of atoms, i.e. the NP volume.\cite{lafiosca2021going}

This work proposes a novel multiscale approach specifically designed to overcome the limitations associated with atomistic and continuum approaches. The method, which is here specified for noble metal NPs, describes the core with an implicit approach, by using the BEM method, while the NP surface is treated at the fully atomistic approach, by means of \wfqfmu. 
The resulting \wfqfmubem approach constitutes, to the best of our knowledge, the first hybrid atomistic-continuum methodology to evaluate the optical response of plasmonic substrates within classical electrodynamics. 
Furthermore, \wfqfmubem is coupled to a QM description of a molecular adsorbate described at the Time-Dependent Density Functional Theory (TDDFT) level, allowing for the calculation of molecular properties and signals in the vicinity of plasmonic nanostructures.\cite{casida1995time,corni2002surface,lafiosca2023classical} 

One of the most interesting aspects of molecular nanoplasmonics is the huge enhancement of the induced electric field in the NP surface proximity, which can drastically affect the electronic properties of molecular adsorbates.  As a result, molecular spectral signals can be significantly enhanced, providing an invaluable platform for molecular sensing. The most diffuse technique that exploits this effect is Surface Enhanced Raman Scattering (SERS),\cite{albrecht1977anomalously,jeanmaire1977surface,campion1998surface, Talley2005Surface,willets2007localized,morton2009understanding,
alvarez2010light,le2008principles,maier2007plasmonics,SHARMA201216,langer2019present,han2022surface}
where the molecular Raman signals are enhanced by several orders of magnitude up to allowing single molecule detection. SERS is nowadays used in various applicative fields, such as catalysis, chemical biology, biophysics, and biomedicine.\cite{lai2018recent,zhang2020core,Su2021Recent,zhao2020branched,Fang2023Surface,Troncoso2024SERS,Perumal2021Towards,Taheri2023Plasmonic} 
For this reason, in this paper, QM/\wfqfmubem is extended to compute SERS signals.

The paper is organized as follows. In the next section, \wfqfmu and BEM methods are briefly recalled, and the novel \wfqfmubem and QM/\wfqfmubem models are presented. After a brief section reporting on the computational details, the methods are validated by computing NPs optical properties and SERS spectra of pyridine with the novel approaches or employing reference fully atomistic methods (\wfqfmu and QM/\wfqfmu). Conclusions and future perspectives end the paper.

\section{Theory}\label{sec:theory}

\begin{figure}[htp!]
        \centering
        \includegraphics[width=0.48\textwidth]{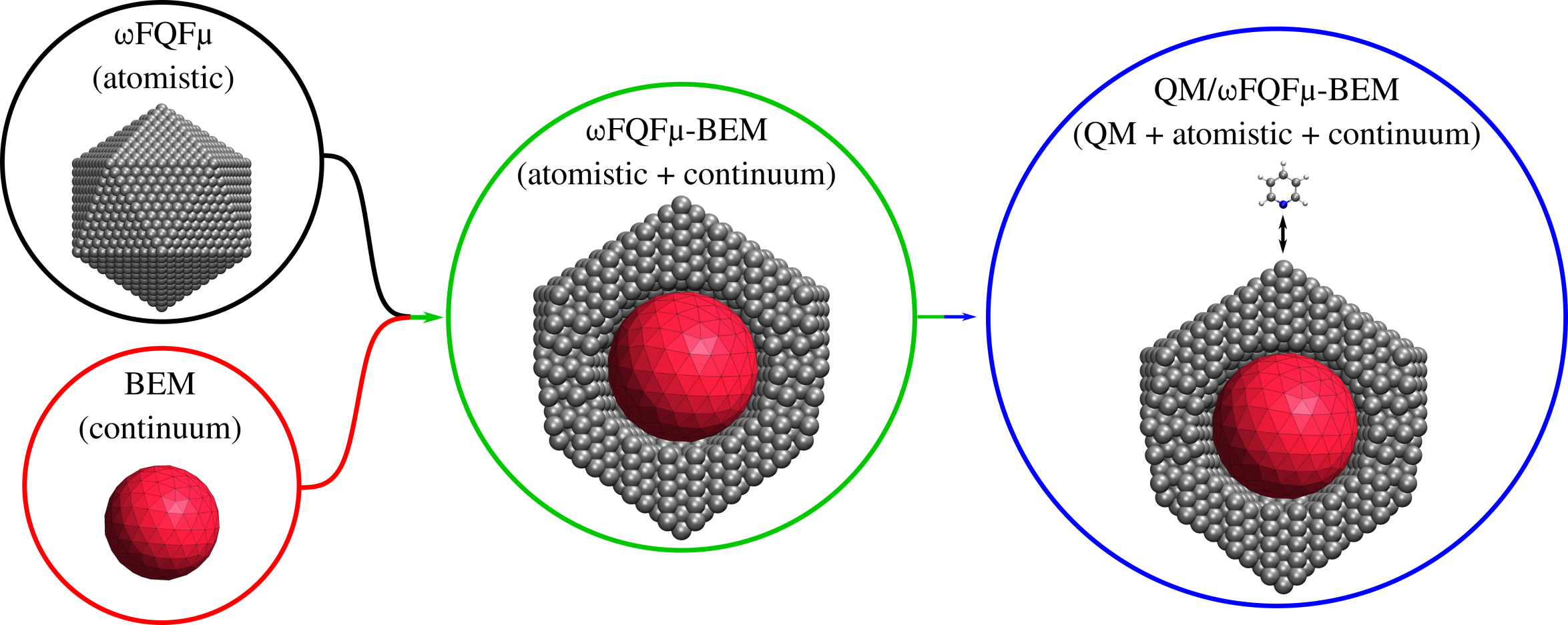}
        \caption{Graphical representation of \wfqfmu, BEM, \wfqfmubem, and QM/\wfqfmubem approaches.} 
        \label{fig:wfqfmubem_scheme}
\end{figure}

This section gives an overview of the theoretical background leading to the formulation of the QM/\wfqfmubem approach. First, the integration of the \wfqfmu and BEM models is discussed, and the \wfqfmubem equations are presented to replicate the plasmonic response of fully atomistic NPs as built by a continuum BEM core within an atomistic \wfqfmu shell (see \cref{fig:wfqfmubem_scheme}). Then, the QM/\wfqfmubem coupling is developed and specified to describe the SERS of a molecule, studied at the TDDFT level, placed in the vicinity of the plasmonic NP.

\subsection{Classical Models for Plasmonics}

\subsubsection*{Frequency-dependent Fluctuating Charges and Fluctuating Dipoles (\texorpdfstring{\wfqfmu}{}) for plasmonic metal nanoparticles}

\wfqfmu is a classical fully atomistic model that accurately reproduces the plasmonic behavior of noble metal nanostructures.\cite{giovannini2019classical,giovannini2022we} In \wfqfmu, each atom is endowed with a complex electric charge ($q$) describing intraband transitions, \cite{giovannini2019classical,bonatti2020plasmonic,giovannini2020graphene,lafiosca2021going,bonatti2022silico} and a complex dipole ($\bm{\mu}$) capturing the interband contributions to the optical response.\cite{giovannini2022we,nicoli2023fully,lafiosca2023classical} Both charges and dipoles are formulated within the quasistatic regime, therefore retardation effects are neglected. Charge exchange occurs through a Drude conduction mechanism\cite{jackson1999classical} confined to nearest neighbor atoms by a damping Fermi function, which mimics the exponential decay of quantum tunneling. \cite{giovannini2019classical,giovannini2022we,nicoli2023fully} For a system composed of $N$ atoms, the Fourier component at frequency $\omega$ of each \wfqfmu charge is computed as:\cite{giovannini2019classical}
\begin{equation}
    -i{\omega}q_i(\omega) = \frac{2n_0\tau}{1-i\omega\tau}\sum_j^N [1-f(r_{ij})]\frac{\mathcal{A}_{ij}}{r_{ij}}(\phi_j^{el}-\phi_i^{el})
    \label{eq:charges}
\end{equation}
where $q_i(\omega)$ is the charge lying on atom $i$ and oscillating at frequency $\omega$, $n_0$ is the atomic density and $\tau$ is the scattering time. $\mathcal{A}_{ij}$ corresponds to the effective area ruling charge exchange between atoms $i$ and $j$, and $r_{ij}$ the distance between these atoms. $f(r_{ij})$ is a phenomenological Fermi damping function parameterized against \textit{ab initio} data.\cite{giovannini2019classical,giovannini2022we,nicoli2023fully} $\phi_i^{el}$ is the electrochemical potential acting on the $i$-th atom, which includes the potential generated by the external electric field, the charges, and the dipoles.

\wfqfmu dipoles add a further source of polarization, that is crucial for modeling the physics of $d$-electrons.\cite{pinchuk2004influence,pinchuk2004optical,balamurugan2005evidence} This is achieved by incorporating a complex frequency-dependent atomic polarizability $\alpha_i^\omega$, extracted from the experimental permittivity, which encloses the effect of interband transitions. The atomic dipoles can then be computed as:\cite{giovannini2022we}
\begin{equation}
    \bm{\mu}_i(\omega) = \alpha^{\omega}_i \textbf{E}^{\text{tot}}_i\label{eq:dipoles}
\end{equation}
where $\bm{\mu}_i(\omega)$ is the complex frequency-dependent atomic dipole on atom $i$. $\textbf{E}_i^{\text{tot}}$ represents the total electric field acting on the $i$-th atomic position, and is the superposition of the external electric field and the electric field generated by the dipoles and charges.

\cref{eq:charges,eq:dipoles} can be reformulated into the following set of linear equations: \cite{giovannini2022we}
\begin{align}
&\sum_{j=1}^N \Bigg(\sum_{k=1}^N \overline{K}_{ik}(\tqq_{kj}-\tqq_{ij}) + \imm\frac{\omega}{w(\omega)}\delta_{ij}\Bigg) q_j \nonumber \\
& + \sum_{j=1}^N\Bigg(\sum_{k=1}^N\overline{K}_{ik}(\tqmu_{kj}-\tqmu_{ij})\Bigg)\bm{\mu}_j \nonumber \\
& = \sum_{k=1}^N\overline{K}_{ik}(\mathbf{V}^\mathrm{ext}_i-\mathbf{V}^\mathrm{ext}_k) \label{eq:q_wfqfmu}\\
& \sum_{j\ne i}^N\tmuq_{ij}q_j + \sum_{j\ne i}^N \tmumu_{ij}\bm{\mu}_j + \frac{1}{\alpha_i^\omega}\bm{\mu}_i = \mathbf{E}_i^\mathrm{ext}  \label{eq:mu_wfqfmu}
\end{align}
where 
$\tqq$, $\tqmu$, and $\tmumu$ are the charge-charge, charge-dipole, and dipole-dipole interaction kernels,\cite{giovannini2019polarizable,giovannini2019classical,giovannini2022we} and $\mathbf{V}_i^{\text{ext}}$ and $\mathbf{E}_i^{\text{ext}}$ are the potential and the electric field generated by the external radiation at each of the $i$-th atomic positions, respectively. In \wfqfmu, the charges and dipoles are associated with a Gaussian-type density distribution, thus the interaction kernels are obtained as first proposed by Mayer.\cite{mayer2007formulation}
Finally, $w(\omega)$ and $\overline{K}_{ij}$ are defined as follows:\cite{lafiosca2021going,giovannini2022we}
\begin{align}
    &{w(\omega)} = \frac{2n_0}{1/\tau -\imm\omega} \label{eq:w_shift} \\ 
    &\overline{K}_{ij} = (1-f(r_{ij}))\frac{\mathcal{A}_{ij}}{r_{ij}} \label{eq:fermi_function}
\end{align}
By defining $\mathbf{T}_{ii}^{\mu\mu} = 1/\alpha^{\omega}$, \cref{eq:q_wfqfmu,eq:mu_wfqfmu} can be recast as: \cite{giovannini2022we}
\begin{equation}
\left[\begin{pmatrix} \mathbf{A}^q & \mathbf{A}^{q\mu} \\ \tmuq & \tmumu\end{pmatrix} - \begin{pmatrix} z(\omega) \mathbf{I}_N & 0 \\ 0 & z'(\omega)\mathbf{I}_{3N} \end{pmatrix}\right]\begin{pmatrix} \mathbf{q} \\ \bm{\mu} \end{pmatrix} = \begin{pmatrix} \mathbf{R} \\ -\mathbf{E}^\mathrm{ext} \end{pmatrix} \label{eq:wfqfmu_final}
\end{equation}
where $\mathbf{A}^q$ represents the charge-charge terms and $\mathbf{A}^{q\mu}$ collects the charge-dipole interactions. The frequency dependence is gathered in the $z$ and $z'$ terms, and $\mathbf{I}_N$ is the $NxN$ identity matrix. Finally, the external potential and external field effects are included in the right-hand-side matrices $\mathbf{R}$ and $\mathbf{E}^\mathrm{ext}$. The reader is referred to section S1 of the Supplementary Material (\sm) for the description of each term in \cref{eq:wfqfmu_final}.
%
%
\subsubsection*{Boundary Element Method (BEM)}

The Boundary Element Method (BEM) constitutes a reliable approach for solving the Polarizable Continuum Model equations applied to the study of plasmonic nanoparticles (PCM-NP).\cite{corni2001theoretical} There, the NP is treated as a homogeneous continuum dielectric by exploiting a classical electrodynamics formalism. Plasmonics is then governed by the frequency-dependent permittivity function chosen to describe the material, and that of the media in which it is embedded.\cite{fuchs1975, trugler2012, bonatti2020plasmonic, mennucci2019multiscale,corni2021role} BEM is applied on a triangular-discretized mesh delimiting the NP surface. At each triangular centroid, a complex point charge that describes the optical response of the material is calculated. 

Here, we rely on the quasistatic PCM-NP framework formulated in the integral equation formalism and numerically solved using BEM. The computation of the BEM charges ($\mathbf{\sigma}$) is described as follows:\cite{corni2001enhanced,tomasi2005quantum,vukovic2009fluorescence,gonzalez2011surface,corni2021role}
\begin{equation}
    \bigg( 2\pi\frac{\varepsilon_2(\omega) + \varepsilon_1(\omega)}{\varepsilon_2(\omega) - \varepsilon_1(\omega)}\mathbf{I}_P + {\textbf{F}}\bigg) \textbf{S}\textbf{A}^{-1}\mathbf{\sigma} = - 
     \big(2\pi\mathbf{I}_P + \mathbf{F}\big)\mathbf{V}
    \label{eq:bem-ief}
\end{equation}
where $\epsilon_{1}(\omega)$ is the complex dielectric function of the environment (for vacuum $\epsilon_{1}(\omega)$=1) and $\epsilon_{2}(\omega)$ that of the NP. $\textbf{A}$ is a diagonal matrix collecting the area of each triangle or $tessera$. $\mathbf{I}_P$ is the $PxP$ identity matrix, where $P$ is the number of $tesserae$. The matrices $\mathbf{S}$ and $\mathbf{F}$ are defined based on the geometrical characteristics of the mesh. $\mathbf{V}$ collects the values of the electrostatic potential generated by the external electric field at each triangular centroid.

We can isolate the frequency-dependent terms of \cref{eq:bem-ief} and reformulate it as follows:
\begin{equation}
    (\mathbf{B}^{\sigma} - \xi(\omega))\mathbf{\sigma} = \mathbf{R_B}
    \label{eq:bem-ief-compact}
\end{equation}

In Section S1 of the \sm, further details on the $\mathbf{F}$ and $\mathbf{S}$ matrices are given, along with the description of each term in \cref{eq:bem-ief-compact}.

%
%
\subsubsection*{\texorpdfstring{\wfqfmubem}{} for plasmonic metal nanoparticles}

The idea of integrating \wfqfmu and BEM, giving rise to a model that is named \wfqfmubem, arises from their close similarities. In particular, both formalisms are grounded in classical electrodynamics and rely on complex frequency-dependent quantities (charges/dipoles) to describe the optical response of metal NPs. However, these models present notable differences. First, the BEM response is rooted in the permittivity chosen to describe the material and lacks any atomistic description. Conversely, \wfqfmu is based on textbook concepts, such as the Drude model of conduction for describing intraband transitions in terms of fluctuating charges. Interband transitions are recovered through a set of fluctuating dipoles, that are defined in terms of the experimental interband polarizability. By this, \wfqfmu can effectively decouple the physical processes influencing the optical response of the plasmonic substrate, while also retaining its atomistic nature. However, \wfqfmu encounters scalability challenges, eventually leading to the exploration of alternative strategies to solve \cref{eq:wfqfmu_final}.\cite{lafiosca2021going,grobas2024multiscale} Remarkably, \wfqfmubem can alleviate these limitations by representing NPs as built of an internal continuum BEM core and an external atomistic \wfqfmu shell, with substantial computational saving.

The \wfqfmubem coupling is formulated by including the potential and electric field generated by the BEM charges in the \wfqfmu equations, specifically in $\phi^{el}$ (\cref{eq:charges}) and $\mathbf{E}^{\text{tot}}$ (\cref{eq:dipoles}), respectively. Additionally, the potential generated by the \wfqfmu charges and dipoles needs to be included in the BEM $\mathbf{V}$ matrix (\cref{eq:bem-ief}). As a result of these modifications, \cref{eq:q_wfqfmu,eq:mu_wfqfmu,eq:bem-ief} are modified as follows:

\begin{align}
&\sum_{j=1}^N \Bigg(\sum_{k=1}^N \overline{K}_{ik}(\tqq_{kj}-\tqq_{ij}) + \imm\frac{\omega}{w(\omega)}\delta_{ij}\Bigg) q_j \nonumber \\
& + \sum_{j=1}^N\Bigg(\sum_{k=1}^N\overline{K}_{ik}(\tqmu_{kj}-\tqmu_{ij})\Bigg)\bm{\mu}_j \nonumber \\
& + \sum_{\nu=1}^P\Bigg(\sum_{k=1}^N\overline{K}_{ik}(\tqsigma_{k\nu}-\tqsigma_{i\nu})\Bigg)\bm{\sigma}_{\nu} \nonumber \\
& = \sum_{k=1}^N\overline{K}_{ik}(\mathbf{V}^\mathrm{ext}_i-\mathbf{V}^\mathrm{ext}_k) \label{eq:q_wfqfmu_bem}\\
\nonumber \\
& \sum_{j\ne i}^N\tmuq_{ij}q_j + \sum_{i=1}^N \sum_{\nu=1}^P \tmusigma_{i\nu}\bm{\sigma}_{\nu} + \sum_{j\ne i}^N \tmumu_{ij}\bm{\mu}_j + \frac{1}{\alpha_i^\omega}\bm{\mu}_i = \mathbf{E}_i^\mathrm{ext}  \label{eq:mu_wfqfmu_bem} \\
\nonumber \\
&\bigg( 2\pi\frac{\varepsilon_2(\omega) + \varepsilon_1(\omega)}{\varepsilon_2(\omega) - \varepsilon_1(\omega)}\mathbf{I}_P + {\textbf{F}}\bigg)(\textbf{S}\textbf{A}^{-1})\mathbf{\sigma} \nonumber \\
&= - 
\big(2\pi\mathbf{I}_P  + \mathbf{F}\big) \big(\mathbf{V}^\mathrm{ext} + \mathbf{V^{{\omega}FQ}} + \mathbf{V^{\omega F\mu}} \big) \label{eq:bem_wfqfmu_bem}
\end{align}
where \cref{eq:q_wfqfmu_bem,eq:mu_wfqfmu_bem} include the effect of the BEM charges ($\mathbf{\sigma}$) in the calculation of the \wfqfmu charges ($\mathbf{q}$) and dipoles ($\bm{\mu}$), respectively. \cref{eq:bem_wfqfmu_bem} extends BEM including the potential generated by the \wfqfmu charges $\mathbf{V^{\omega FQ}}=\tsigmaq_{{\nu}i}q_{i}$ and dipoles $\mathbf{V^{F\mu}} = \tsigmamu_{{\nu}i} \mu_{i}$. $\tsigmaq$ and $\tsigmamu$ are the interaction kernels of BEM point charges with \wfqfmu charges and dipoles described by Gaussian functions, respectively (see Section S1 of the \sm for more details).\cite{molecularelectronicstructuretheory-ch9,giovannini2019classical,giovannini2022we} \\

Finally, the linear problem represented by equations \cref{eq:q_wfqfmu_bem,eq:mu_wfqfmu_bem,eq:bem_wfqfmu_bem} can be expressed in compact matrix notation as follows:

\begin{widetext}
\begin{tikzpicture}[overlay, remember picture, black, thick]
    \draw ([yshift=-3pt]                         pic cs:wfqfmu) -- ++(2.20,0.00); 
    \draw ([xshift=90pt, yshift=-3pt]            pic cs:wfqfmu) -- ++(3.40,0.00); 
    \draw ([xshift=208pt,yshift=-3.3pt]            pic cs:wfqfmu) -- ++(0.43,0.00); 
    \draw ([xshift=246pt,yshift=-3pt]            pic cs:wfqfmu) -- ++(0.95,0.00); 
    \draw ([xshift=43pt,yshift=-15pt]            pic cs:wfqfmu) -- ++(0.00,1.23); 
    \draw ([xshift=161pt,yshift=-15pt]           pic cs:wfqfmu) -- ++(0.00,1.23); 
\end{tikzpicture}
\begin{equation}
\left[\begin{pmatrix}
\mathbf{A}^q                             &    \mathbf{A}^{q \mu}           &    \mathbf{A}^{q\sigma}       \\
\tikzmark{wfqfmu}\mathbf{T}^{\mu q}                       &    \mathbf{T}^{\mu \mu}         &    \mathbf{T}^{\mu \sigma}    \\
\mathbf{B}^{\sigma q}                    &    \mathbf{B}^{\sigma \mu}      &    \mathbf{B}^{\sigma}               \\
\end{pmatrix}
-
\begin{pmatrix}
  z(\omega)\mathbf{I}_N         & 0                 &   0               \\
  0                    & z'(\omega)\mathbf{I}_N     &   0               \\
  0                    & 0                 &   \xi(\omega)  \\
\end{pmatrix}
\right]
\begin{pmatrix} 
\mathbf{q} \\ \mathbf{\mu} \\ \mathbf{\sigma} 
\end{pmatrix}
=
\begin{pmatrix}
\mathbf{R} \\  -\mathbf{E}^\mathrm{ext}  \\  \mathbf{R_{B}} 
\end{pmatrix}
\label{eq:final_wfqfmu_bem}
\end{equation}
\end{widetext}

The terms in \cref{eq:final_wfqfmu_bem} are detailed in Section S1 of the SI.

\subsection{QM/\wfqfmubem model and its extension to SERS}\label{theoqmm}

Once \wfqfmubem has been formulated, it can be coupled to a QM description of a molecular system in a QM/MM fashion,\cite{warshel1976theoretical,lin2007qm,senn2009qm,mennucci2019multiscale,morton2010discrete,guido2020open,coccia2020hybrid,corni2015equation} giving rise to the QM/\wfqfmubem model. \wfqfmubem describes the system's response to external oscillating electric fields, therefore it can be naturally translated into a linear response formalism. The coupling is done by following the same strategy that was exploited to formulate the two-layer QM/\wfqfmu approach. \cite{lafiosca2023classical}.


Linear response equations in the Time-Dependent Kohn-Sham (TDKS) framework read:\cite{casida1995time,corni2002surface,norman2018principles}
\begin{equation}
   \begin{bmatrix}
   \begin{pmatrix} 
        \mathbf{A}      & \mathbf{B} \\ 
        \mathbf{B}^\ast & \mathbf{A}^\ast
    \end{pmatrix}    
   -(\omega + \text{i}\Gamma)
   \begin{pmatrix}
       \mathbf{I} & 0\\
       0 & -\mathbf{I}
   \end{pmatrix}
   \end{bmatrix}
   \begin{pmatrix}
       \textbf{X}\\
       \textbf{Y}
   \end{pmatrix}
   = - 
    \begin{pmatrix}
       \textbf{Q}\\
       \textbf{Q}\ast
    \end{pmatrix}
    \label{eq:cpks}
\end{equation}

In the QM/\wfqfmubem approach, the terms entering the linear system in eq.\ref{eq:cpks} are modified to account for the presence of the \wfqfmu and BEM layers, i.e.

\begin{equation}\label{eq:defs-cpks}
\begin{aligned}
A_{ai,bj} & = (\epsilon_a - \epsilon_i)\delta_{ab}\delta_{ij} + (ai|bj) - c_x (ab|ij) + c_l f^{xc}_{ai,bj} + \\
&\hspace{0.5cm}C^{\text{QM}/\omega\text{FQF}\mu\text{-BEM}}_{ai,bj} \\
B_{ai,bj} & = (ai|bj) - c_x(aj|ib) + C^{\text{QM}/\omega\text{FQF}\mu\text{-BEM}}_{ai,bj} \\
Q_{ia} & = \braket{\phi_i|V^{\alpha,\mathrm{pert}}(\mathrm{r},\omega)|\phi_a}
\end{aligned}
\end{equation}

In \cref{eq:cpks}, $\mathbf{X}$ and $\mathbf{Y}$ are the excitation and de-excitation transition densities, respectively, while $\Gamma$ is a phenomenological damping factor. Furthermore, in \cref{eq:defs-cpks}, the $(i,j)$/$(a,b)$ indices run over occupied/virtual Kohn-Sham molecular orbitals. $\varepsilon$ is the molecular orbital energy, ($ai|bj$) denotes two-electron integrals, and the $c_x,c_l$ parameters vary depending on the DFT functional employed. Additionally, $\mathbf{C}^{\text{QM}/\omega\text{FQF}\mu\text{-BEM}}$ describes the polarization induced by the \wfqfmubem charges and dipoles, which respond to the perturbed TDKS density. On the right-hand-side, $Q_{ia} = \braket{\phi_i|V^{\alpha,\mathrm{pert}}(\mathrm{r},\omega)|\phi_a}$ represents the perturbation operator polarized along $\alpha$. This operator includes the potential generated by the external electric field and the local field operator. The latter captures the polarization generated due to the plasmonic behavior of the substrate, reacting to the external electric field. 

In the current implementation in the AMS Software,\cite{ams2020} the Kohn-Sham operators are generated on a grid of points exploiting a numerical integration strategy. In this framework, the polarization induced by the \wfqfmubem charges and dipoles ($\mathbf{K}^{\text{pol}}$), calculated as a function of the distance between the $l$-th atom or $tessera$ centroid and a generic grid point ($d_l$), is given by:
\begin{equation}\label{eq:polarization terms}
\begin{aligned}
    \mathbf{K}^{\text{pol}}(\mathbf{r},\omega) = &\sum_{l=1}^N q_l(\omega) \mathbf{T}^{(0)}(d_l) + &\sum_{l=1}^N\bm{\mu}_l(\omega)\cdot\mathbf{T}^{(1)}(d_l) + \\ 
    &\sum_{l=1}^P {\sigma}_l(\omega) \mathbf{T}^{(0)}(_l)
\end{aligned}
\end{equation}
where $\mathbf{T}^{(0)}$ and $\mathbf{T}^{(1)}$ are the charge-to-grid and dipole-to-grid interaction kernels, respectively.\cite{nicoli2022assessing,lafiosca2023classical}
QM/\wfqfmubem can be extended to compute SERS spectra of molecular systems in the vicinity of metal NPs. To this end, the complex polarizability tensor $\Bar{\alpha}_{\alpha\beta}$ needs to be computed,\cite{lafiosca2023classical,jensen2005finite} where $\alpha/\beta$ represent the Cartesian components ($x,y,z$). After solving \cref{eq:cpks}$, \Bar{\alpha}_{\alpha\beta}$ can be calculated as follows:
\begin{equation}\label{eq:polar}
\bar{\alpha}_{\alpha\beta}(\omega;\omega') = -\mathrm{tr}\left[\mathbf{H}^\alpha(\omega)\mathbf{P}^\beta(\omega')\right],
\end{equation}
where $\mathbf{H}^\alpha(\omega)$ is the dipole matrix, which includes the QM dipole and the \wfqfmubem local field operators polarized along $\alpha$, while $\mathbf{P}^{\beta}(\omega')$ is the first-order perturbed density matrix.\cite{corni2001enhanced,payton2014hybrid,morton2011discrete,lafiosca2023classical}

According to Placzek's theory of Raman scattering,\cite{placzek1934handbuch} the Raman signal can be modeled from the frequency-dependent polarizability tensor. Assuming the frequency of the incident and scattered fields to coincide, the Raman intensity related to the $k$-th normal mode $I^k$ of the QM molecule is given by:\cite{placzek1933rotationsstruktur,placzek1934handbuch,jensen2005theory,lafiosca2023classical,corni2006studying}
\begin{equation}\label{eq:raman-int-full}
I^k \propto \frac{(\omega - \omega_k)^4}{\omega_k}\{45[\alpha_k'(\omega-\omega_k;\omega)]^2 + 7[\gamma_k'(\omega-\omega_k;\omega)]^2
\end{equation}
where $\omega$ is the incident frequency, while $\omega_k$ and $\alpha_k'/\gamma_k'$, are the frequency and the isotropic/anisotropic polarizability derivatives associated with the $k$-th normal mode, respectively.

\section{Computational Details}\label{sec:computation_details}

Ag structures are created by using the Atomic Simulation Environment (ASE) Python module v. 3.17.\cite{larsen2017atomic} Specifically, Ag atoms are disposed in a Face-Centered Cubic (FCC) arrangement defined by a lattice parameter of 4.08 \AA. Two morphologies are studied: spherical and icosahedral (Ih) NPs. For \wfqfmu calculations, the parameters defined in Ref. \citenum{giovannini2022we} are exploited. Detailed information regarding the geometrical characteristics of all atomistic structures are reported in Tables S1--S4 of the \sm. 

The continuum BEM meshes representing the surface of both spherical and Ih NPs are constructed by using the GMSH software.\cite{gmsh} The frequency-dependent permittivities proposed by Palik,\cite{palik1997handbook} Brendel-Bormann,\cite{aleksandar1998laser} and Johnson and Christy\cite{johnson1972optical} are exploited in BEM calculations.

\wfqfmubem spherical and Ih structures are built as a core, described as an implicit spherical NP and treated at the BEM level, surrounded by an atomistic shell, treated at the \wfqfmu level. To validate the novel approach, we investigate the variation of the optical response of a spherical Ag nanostructure (radius 4 nm) as a function of the method parameters, namely the core-to-shell distance, the thickness of the atomistic shell, the BEM \emph{tessellation}, and the dielectric function to describe BEM response. 

In particular, we study the optical response as a function of the minimum core-to-shell distance, which varies from 2.88 \AA~ to 5.76 \AA~, corresponding to integer multiples of the nearest neighbor distance (2.88 \AA) in the studied FCC lattices. For Ag Ih structures, the atomistic shell is defined ensuring a minimum thickness of 3 atomic layers in the thinner region. 

In \wfqfmubem, the total complex dipole of the NP ($\Bar{\zeta}$) is computed as:
\begin{equation*}
      \bar{\zeta} (\omega)= \sum_i^N q_i (\omega) \mathbf{r}_i+\sum_i^N \bm{\mu}_i (\omega) + \sum_\nu^P \sigma_\nu (\omega) \mathbf{s}_\nu 
\end{equation*}
where $\mathbf{r}_i$ and $\mathbf{s}_{\nu}$ are the position of the $i$-th atom and of the $\nu$-th $tessera$ centroid, respectively. From $\Bar{\zeta}$, the NP complex frequency-dependent polarizability ($\Bar{\xi}$) and the absorption cross section ($\sigma^{\text{abs}}$) can be calculated as:
\begin{equation}
    \bar{\xi}(\omega)_{\alpha\beta} = \frac{\bar{\zeta}_\alpha}{E_{0,\beta}}  \quad \Longrightarrow \quad
    \sigma^{\text{abs}}(\omega) = \dfrac{4\pi\omega}{3c} \text{Tr} (\bar{\xi})
\end{equation}
where $\alpha\beta$ indicates the polarization of the incident electric field ($E_0$), and $c$ is the speed of light.

After validating \wfqfmubem, we exploit the multiscale QM/\wfqfmubem to calculate SERS signals of pyridine adsorbed on Ag Ih NPs (with radius varying from 1.9 to 3.8 nm). SERS spectra are computed by representing the Ag Ih NP at the fully atomistic \wfqfmu, implicit BEM, and hybrid \wfqfmubem levels, thus allowing for validating the novel method by a robust comparison with state-of-the-art methods. In all cases, pyridine is placed longitudinally at a distance of 3 \AA~ from the NP surface. Such distance is calculated from the nitrogen atom to the nearest NP atom/$tessera$ centroid defining the tip of the atomistic/continuum surface of Ag Ih structures. 

In all calculations, pyridine is described at the QM level utilizing the BP86 functional and a double-$\zeta$-polarized DZP basis set, in agreement with previous studies.\cite{payton2014hybrid,lafiosca2023classical} TDDFT equations are solved by imposing a damping factor $\Gamma = 0.01$ eV.\cite{van1995density,van1999implementation,payton2014hybrid,lafiosca2023classical} $\alpha_k'$ and $\gamma_k'$ (see \cref{eq:raman-int-full}) are calculated by a numerical differentiation scheme, with a constant step size of 0.001 \AA.\cite{lafiosca2023classical,fan1992application,fan1992application2,van1996application} Normal modes' displacements of pyridine are evaluated \textit{in vacuo}, without explicitly considering NP effects, which are expected to be small for the considered system.\cite{lafiosca2023classical,payton2014hybrid} The influence of the plasmon resonance on SERS signals is assessed by matching the incident external Raman frequency with the PRF of the NP. Finally, SERS spectra are plotted using Lorentzian band-shapes characterized by a full width at half maximum (FWHM) of 4 cm$^{-1}$. QM/\wfqfmubem, QM/\wfqfmu, and QM/BEM calculations are performed by using a locally modified version of the Amsterdam Modelling Suite (AMS).\cite{ams2020} 

The NP effect on Raman signals is evaluated through the definition of three observables: the Enhancement Factor (EF), the Maximum Enhancement Factor (MEF), and the Averaged Enhancement Factor (AEF), which are defined as follows:
\begin{equation} \label{eq:ef}
    \mathrm{EF}^k(\omega) = \frac{I^k_\mathrm{NP}(\omega)}{I^k_\mathrm{vac}(\omega)}    
\end{equation}
\begin{equation} \label{eq:mef}
    \mathrm{MEF}(\omega)  = \max_k \mathrm{EF}^k(\omega)    
\end{equation}
\begin{equation} \label{eq:aef}
    \mathrm{AEF}(\omega)  = \frac{\sum_k I^k_\mathrm{NP}(\omega)}{\sum_l I^l_\mathrm{vac}(\omega)}    
\end{equation}
where $I_{\text{NP}}^k$ and $I_{\text{vac}}^k$ are the molecular Raman intensities associated with the $k$-th normal mode (see \cref{eq:raman-int-full}) in the presence of NP and \textit{in vacuo}, respectively. In \cref{eq:aef}, $k$ and $l$ indices run over the set of studied molecular normal modes.

\section{Results and Discussion}\label{sec:results}

In this section, \wfqfmubem and QM/\wfqfmubem methods are validated by comparison with the reference \wfqfmu and QM/\wfqfmu approaches. In particular, \wfqfmubem is first challenged to reproduce the plasmonic features of spherical NPs, by analyzing how a variation of the model parameters affects the optical response. Then, QM/\wfqfmubem is applied to compute SERS of pyridine adsorbed on complex-shaped Ih nanostructures.

\subsection{The \texorpdfstring{\wfqfmubem}{} method for Plasmonics}

\subsubsection*{Plasmonic Features of Single Nanoparticles}

\begin{figure}[htp!]
        \centering
        \includegraphics[width=0.48\textwidth]{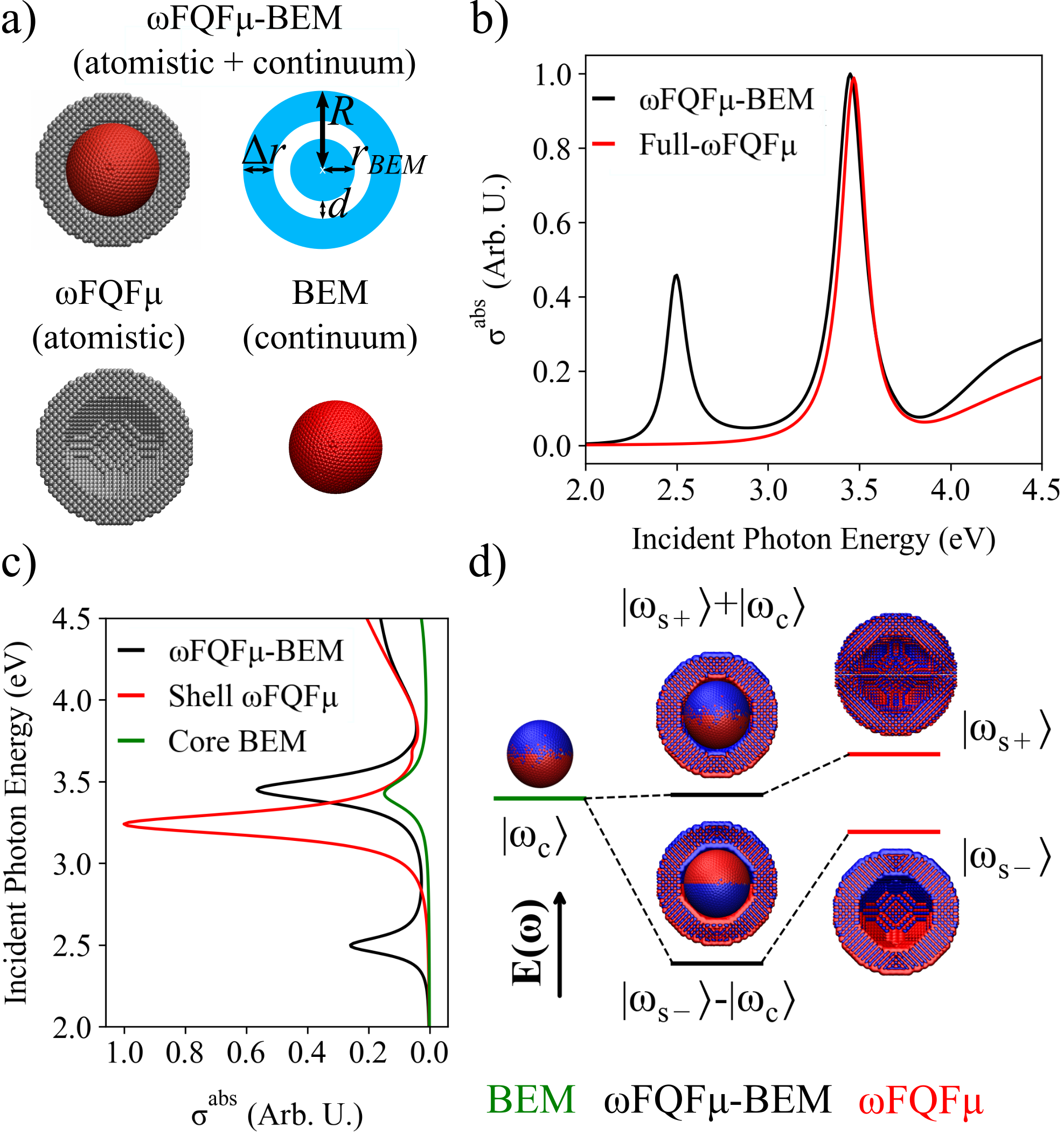}
        \caption{a) Graphical representation and geometrical parameters of a spherical NP described by \wfqfmubem. b) Normalized absorption cross-sections $\sigma^{\text{abs}}$ of a spherical Ag NP (radius = 40 \AA) computed at the \wfqfmubem and full-\wfqfmu levels. c) Normalized absorption spectra of the selected spherical NP calculated at the \wfqfmubem level, together with the absorption spectra of the BEM core and \wfqfmu shell. (see panel a). d) Schematic picture of the hybridization of BEM charge distribution (left), and \wfqfmu densities (right), leading to \wfqfmubem charges and densities (center) calculated at the corresponding PRFs.}  
        \label{fig:plasmonic-modes-d5.76-bb}
\end{figure}

In \cref{fig:plasmonic-modes-d5.76-bb}a, we show a graphical representation of a spherical NP described at the \wfqfmubem level, constructed by integrating a continuum BEM spherical core within an atomistic \wfqfmu spherical shell. \wfqfmubem spherical NP geometries are defined by four parameters: the radius of the inner BEM core ($r_{\text{BEM}}$), the radii of the outer \wfqfmu shell ($R$), the BEM core - \wfqfmu shell distance ($d$), and the difference between the radii of the inner and outer shells ($\Delta r$). The computed optical response depends on the proper definition of such parameters, the dielectric function exploited to model the BEM portion, and the number of $tesserae$ exploited to mesh the BEM region. It is worth noting that our approach is general and the full BEM ($r_{\text{BEM}} = R$) or the full \wfqfmu ($\Delta r = R$) description can be easily recovered. 

Let us first focus on the computed response for a specific geometry, 
i.e a spherical NP with $R=40.00$ \AA, $r_{\text{BEM}}=20.26$ \AA, $\Delta r =13.98$ \AA, and $d=5.76$ \AA~. 11470 atoms constitute the atomistic shell, while the BEM core is defined by 2972 \textit{tesserae}. The Brendel-Bormann frequency-dependent permittivity\cite{aleksandar1998laser} is exploited to describe the BEM part. 

\Cref{fig:plasmonic-modes-d5.76-bb}b compares normalized absorption spectra calculated for \wfqfmubem and the reference, fully atomistic \wfqfmu (hereafter referred to as {\it{full-\wfqfmu}}). The normalization is performed with respect to the full-\wfqfmu PRF absorption. The full-\wfqfmu approach predicts a single absorption peak at 3.47 eV, whereas \wfqfmubem spectrum features two peaks centered at 2.50 eV and 3.45 eV. To analyze the physical origin of such a discrepancy, in \cref{fig:plasmonic-modes-d5.76-bb}c normalized absorption spectra of the selected spherical NP calculated at the \wfqfmubem level are reported, together with the absorption spectrum of the BEM core and \wfqfmu shell. Spectra are normalized with respect to the maximum absorption of the \wfqfmu shell. The spherical shell spectrum shows a high-intensity peak centered at 3.24 eV and a low-intensity peak at 3.68 eV. These two peaks can be assigned to the bonding ($\ket{\omega_{\text{s+}}}$) and anti-bonding ($\ket{\omega_{\text{s-}}}$) plasmonic modes, respectively (see \cref{fig:plasmonic-modes-d5.76-bb}d), that are typical of plasmonic nanoshells.\cite{park2009nature,bardhan2010nanosphere,halas2011plasmons,bonatti2020plasmonic} A single plasmonic peak at about 3.43 eV ($\ket{\omega_{\text{c}}}$) is instead present in the spectrum of the BEM core. Such a band corresponds to the dipolar plasmonic excitation (see \cref{fig:plasmonic-modes-d5.76-bb}d). 

The two peaks reported in the \wfqfmubem spectrum of the spherical NP are thus due to the hybridization of isolated plasmonic modes from the core and shell, similar to what has been reported for other nanostructures.\cite{prodan2003hybridization,nordlander2004plasmon,halas2011plasmons,nordlander2013quantum,marinica2016quantum} More specifically, \wfqfmubem peak at 2.50 eV is associated with the hybrid $\ket{\omega_{\text{s-}}}-\ket{\omega_{\text{c}}}$ mode, while the band centered at 3.45 eV corresponds to the $\ket{\omega_{\text{s+}}}+\ket{\omega_{\text{c}}}$ mode combination (see \cref{fig:plasmonic-modes-d5.76-bb}d). The presence of the low-energy peak is not unexpected because the approach discards any charge transfer between the two classical regions. This peak thus arises as an artifact of our model. However, it can be easily identified because it is the lowest energy peak computed in the spectrum, as predicted by hybridization theory. In the following, the focus will thus shift to the $\ket{\omega_{\text{s+}}}+\ket{\omega_{\text{c}}}$ band, whose PRF remarkably aligns with that predicted at the full-\wfqfmu level.



\subsubsection*{Plasmonic response dependence on the thickness of the atomistic shell}\label{sec:thickness-dependence}

This section first examines how a variation in the shell thickness and core-shell distance $d$ influence the optical response of the metal NPs. Specifically, the shell thickness $\Delta r$ varies from 13.98 to 7.93 \AA, with a constant step of $\sim$2.0 \AA. Two core-to-shell distances are considered $d=2.88$ \AA~ and $d=5.76$ \AA~, corresponding to integer multiples of the nearest neighbor distance (2.88 \AA) in the studied FCC lattice. In all cases, the BEM matrix is described by the Brendel-Bormann dielectric function.\cite{aleksandar1998laser} 

\begin{figure}[htb!]
        \centering
        \includegraphics[width=0.48\textwidth]{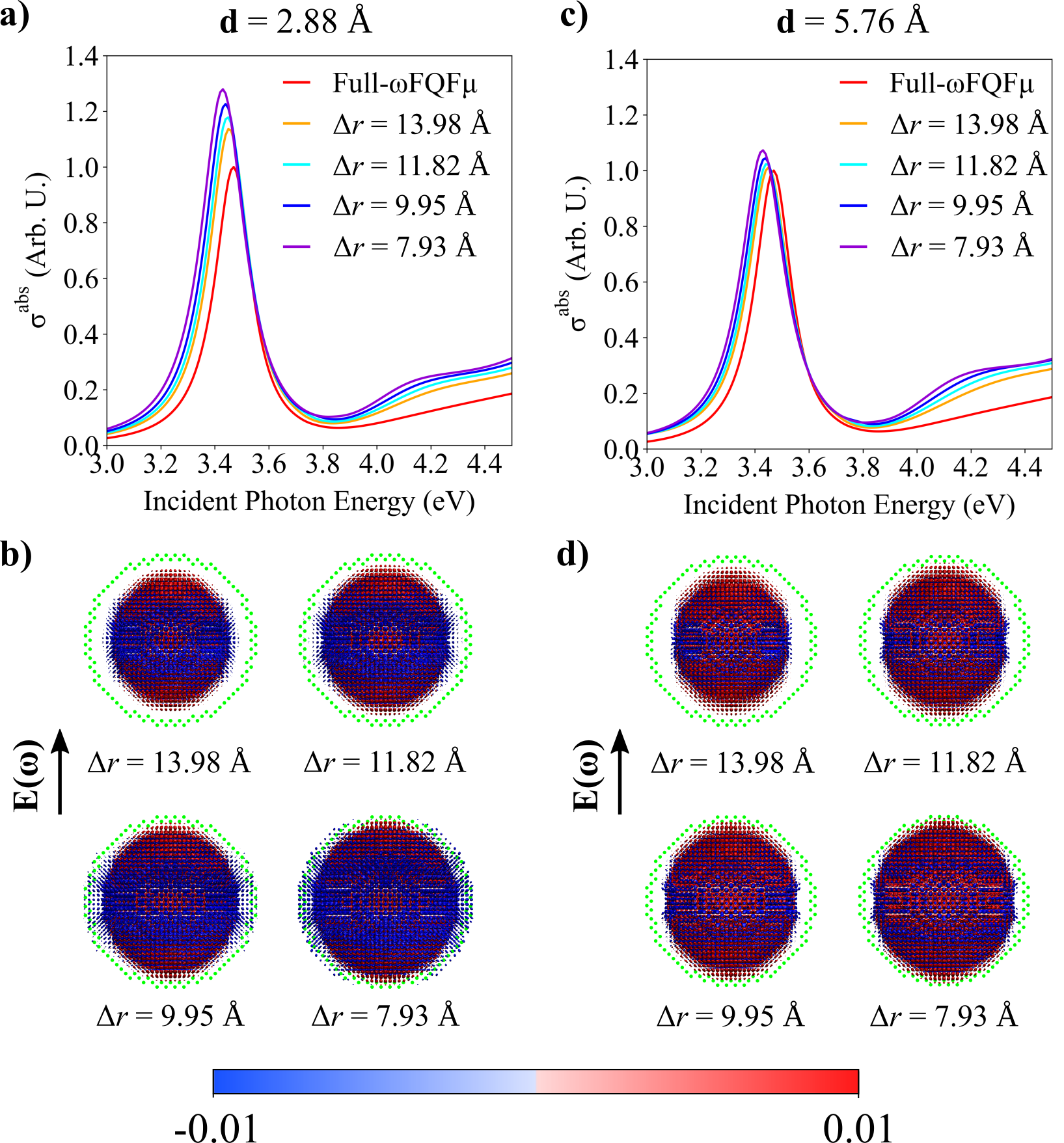}
        \caption{Normalized absorption spectra of a spherical NP (a,c) and absolute density differences plots calculated at the PRFs (b,d - green dots represent surface NP atoms.) as a function of $\Delta r$. The NP is described at the full-\wfqfmu and \wfqfmubem levels, setting $d=5.76$ \AA~(a,b) and $d=2.88$ \AA~(c,d).} 
        \label{fig:thickness-dependence-bb}
\end{figure}

\cref{fig:thickness-dependence-bb} (a,c) reports absorption cross-sections of the spherical NP as a function of shell thickness $\Delta r$ for $d = 2.88$ and $d = 5.76$ \AA, respectively. Data are normalized with respect to the full-\wfqfmu PRF absorption. As a reference, the normalized absorption spectrum computed at the full-\wfqfmu level is reported in all panels. All \wfqfmubem spectra feature an intense peak in the region 3.0-4.5 eV. Such a peak is associated with a dipolar plasmon, as commented in the previous section. \wfqfmubem computed PRFs remain almost constant as a function of the shell thickness, and only present slight shifts of 0.02 to 0.04 eV with respect to \wfqfmu PRF (3.47 eV), regardless of the value of $d$. Differently, reducing the shell thickness systematically increases the absorption intensity mismatch at the PRF between \wfqfmubem and the reference \wfqfmu data. Notably, the mismatch becomes more pronounced for structures with $d=2.88$ \AA. This behavior is probably due to numerical instabilities occurring when solving the \wfqfmubem linear equation (see \cref{eq:final_wfqfmu_bem}) for the smaller $d$ values, associated with the so-called polarization catastrophe raising for the increase of electrostatic/polarization interactions.\cite{thole1981molecular} 

To further analyze the plasmonic response as a function of $d$, \cref{fig:thickness-dependence-bb} (b,d) graphically illustrates the differences in the computed plasmonic density at the \wfqfmubem and full-\wfqfmu levels. In all cases, densities are computed at the PRF for $d = 2.88$ \AA~ and $d=5.76$ \AA, respectively. This qualitative analysis permits us to evaluate the plasmon-density deviation near the NP surface, which is graphically represented by green dots. The most remarkable differences are observed at the core region, and in particular at the BEM-\wfqfmu interface. This is expected and is due to the artificial boundary that is introduced in our multiscale method. However, since we are dealing with localized surface plasmons, we are particularly interested in a proper description of surface properties. Note that the differences at the NP surface are negligible for all methods, thus validating the novel methodology. The agreement between the two methods worsens by reducing the atomistic shell thickness, especially for $d = 2.88$ \AA. This is again related to numerical instabilities in solving the \wfqfmubem linear equation. 

For a more quantitative analysis of the plasmonic density differences close to the NP, in \cref{table:thickness-dependence-d5.76} computed relative errors ($\rho^{\text{error}}$) between integrated densities of \wfqfmubem and full-\wfqfmu are collected.  Values are calculated at their corresponding PRFs. $\rho^{\text{error}}$ is evaluated within a volume $V$ close to the NP surface, defined as a three-dimensional parallelogram constrained by the coordinate ranges $x, z \in [-10, 10]$ \AA\ and $y \in [40, 50]$ \AA~ (see Fig. S1 of the \sm for a graphical representation). $\rho^{\text{error}}$ is calculated as follows:
\begin{equation}
    \rho^{\text{error}} = \frac{\int_V|\rho_{\text{PRF}}^{\omega\text{FQF}\mu\text{-BEM}} (\mathbf{r})- \rho_{\text{PRF}}^{\text{Full-}\omega\text{FQF}\mu}(\mathbf{r})| \text{d}\mathbf{r^3}}{\int_V|\rho_{\text{PRF}}^{\text{Full-}\omega\text{FQF}\mu}(\mathbf{r})|\text{d}\mathbf{r}}\cdot 100
\end{equation}
where $\rho_{\text{PRF}}^{\omega\text{FQF}\mu\text{-BEM}}$ and $\rho_{\text{PRF}}^{\text{Full-}\omega\text{FQF}\mu}$ are the \wfqfmubem and full-\wfqfmu densities calculated at their PRF, respectively. The results reported in \cref{table:thickness-dependence-d5.76} confirm the qualitative behavior depicted in \cref{fig:thickness-dependence-bb}. 
In fact, by decreasing the shell thickness, the computed $\rho^{\text{error}}$ values increase for both $d = 2.88$ \AA~(ranging from about 12\% to about 21\%) and $d = 5.76$ \AA~ (ranging from about 3\% to about 10\%). Also, this quantitative analysis confirms the better numerical performance of the \wfqfmubem structures with $d=5.76$ \AA, which are consistently associated to the lowest relative errors. Remarkably, for the largest $\Delta r$, \wfqfmubem is associated with a relative error of about 3\%, thus validating our novel approach as compared to a full atomistic description.


\begin{table}[htbp!]
    \centering
    \begin{tabular}{cccccc}
    \toprule
        $\Delta r$ (\AA)       &N$_{\text{Atoms}}$           & $r_{\text{BEM}}$ (\AA)         & PRF (eV)          & Speed-up (\%) &  $\rho^{\text{error}}$(\%) \\ \midrule
                        & \multicolumn{5}{c}{{$d$}=2.88\,\r{A}} \\ \cmidrule(r{2pt}){2-6}
        13.98            & 11470            & 23.14              & 3.45              &  44.35        & 11.74  \\
        11.82            & 10258            & 25.30              & 3.45              &  56.87        & 14.36  \\
         9.95            &  9080            & 27.17              & 3.44              &  66.01        & 19.34  \\
         7.93            &  7676            & 29.19              & 3.43              &  77.32        & 21.25  \\
                         &                   &               &       \\
                        & \multicolumn{5}{c}{{$d$}=5.76\,\r{A}} \\ \cmidrule(r{2pt}){2-6}                  
        13.98            & 11470            & 20.26              & 3.45              &  42.56        & 3.49   \\
        11.82            & 10258            & 22.42              & 3.44              &  53.55        & 5.45   \\
         9.95            &  9080            & 24.29              & 3.44              &  65.20        & 6.77   \\
         7.93            &  7676            & 26.31              & 3.43              &  75.81        & 10.42  \\
                         &                  &                    &                   &               & \\
        Full-\wfqfmu     & 15683            &  --                & 3.47              &  --           & --     \\ \bottomrule
    \end{tabular}
    \caption{PRFs of the studied spherical NP ($R = 40$ \AA) calculated at the \wfqfmubem level (${d}=2.88$ \AA~ and $d=5.76$ \AA) as a function of geometrical parameters ($\Delta r$, $N_{\text{{Atoms}}}$, $r_{\text{BEM}}$). \wfqfmubem computational \% speed-up and $\rho^{\text{error}}$ with respect to full-\wfqfmu reference are also given.}
    \label{table:thickness-dependence-d5.76}
\end{table}

To conclude this section, we investigate the computational savings associated with the multiscale \wfqfmubem as compared to a full \wfqfmu description of the spherical NP. To this end, the relative computational speed-up (in percentage) for single-frequency calculations (at the PRF of the systems) are reported in \cref{table:thickness-dependence-d5.76} as a function of $d$ and $\Delta r$. We remark that all calculations exploit a constant number of \emph{tesserae} to mesh the BEM core ($\sim$ 2980 \textit{tesserae}). The data reported in \cref{table:thickness-dependence-d5.76} clearly show a substantial computational saving, ranging from about 43\% to about 77\%, independently of $d$ for a given value or $\Delta r$. This is expected because the number of atoms in the \wfqfmu shell decreases while the number of \emph{tesserae} representing the BEM core remains constant. It is worth remarking that for the most accurate \wfqfmubem partitioning ($\Delta r = 13.98$ \AA, $d = 5.76$ \AA) the speed-up of \wfqfmubem is considerable, without losing accuracy as compared to a full \wfqfmu description. Therefore, the above model parameters represent the best compromise between accuracy and computational cost for \wfqfmubem applications.


\subsubsection*{Plasmonic response dependence on the BEM dielectric functions}\label{sec:dielectric-comparison}


\begin{figure*}[htp!]
        \centering
        \includegraphics[width=\textwidth]{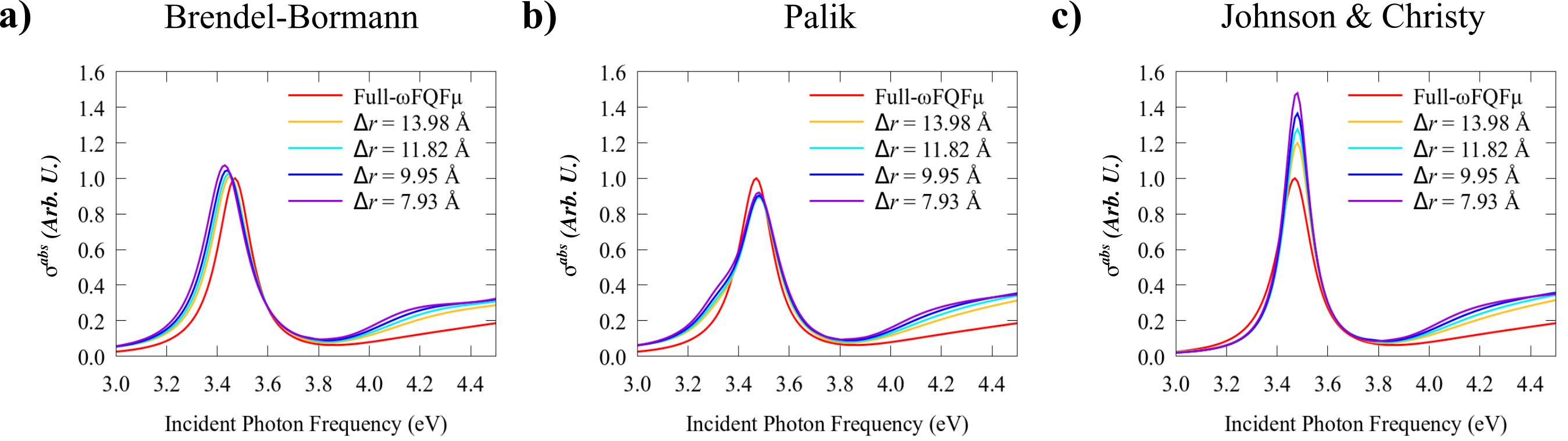}
        \caption{Normalized absorption spectra of a spherical Ag NP ($R = 40$ \AA) treated at the \wfqfmubem ($d=5.76$ \AA) levels as a function of $\Delta r$ and BEM dielectric function [Brendel-Bormann (a), Palik (b), and Johnson and Christy (c)]. Full-\wfqfmu spectra are also depicted as a reference.}
        \label{fig:thickness-epsilon-dependence-d5.76}
\end{figure*}

We now move to analyze the model parameters affecting the BEM core response, namely the dielectric function and the number of \emph{tesserae}. To this end, we consider three dielectric functions that are commonly exploited to describe Ag response, recovered from Brendel-Bormann (BB),\cite{aleksandar1998laser} Palik,\cite{palik1997handbook} and Johnson and Christy (J\&C).\cite{johnson1972optical} 
Computed absorption cross sections for the selected spherical NP exploiting the three permittivity functions are normalized with respect to the full-\wfqfmu reference and graphically depicted in \cref{fig:thickness-epsilon-dependence-d5.76} (a-c), for $d = 5.76$ \AA~ (see Fig. S2 of the \sm for additional values obtained with $d = 2.88$ \AA~). Absorption properties are reported as a function of the shell thickness $\Delta r$. The normalized full atomistic \wfqfmu absorption spectrum is also shown as a reference. 

All computed spectra are characterized by a main plasmonic peak, which is associated with a dipolar localized surface plasmon. \wfqfmubem computed PRFs (see also \cref{table:prf-rho-dielectric}) remain constant as a function of the shell thickness by exploiting Palik and J\&C permittivity functions (3.48 eV), and remarkably the computed PRF is shifted by only 0.01 eV with respect to the reference \wfqfmu PRF (3.47 eV). As commented in the previous section, by using the BB $\varepsilon(\omega)$, the PRF slightly shifts by decreasing the shell thickness (from 3.45 eV - $\Delta r = 13.98$ \AA~ - to 3.43 eV - $\Delta r = 7.93$ \AA). However, also in this case all data agree with the reference \wfqfmu PRF within the chemical accuracy (0.04 eV $\sim$ 0.9 kcal/mol).

The absorption cross-section at the PRF varies significantly depending on the chosen dielectric function. Notably, absorption intensities in very good agreement with the reference \wfqfmu model are obtained by employing BB and Palik dielectric functions. The \wfqfmubem spectrum computed by using the Palik dielectric function presents a small shoulder at about 3.3 eV, which is also observed at the full BEM level (see Fig. S3 in the \sm), and is thus related to the use of this specific permittivity function. In contrast, a pronounced intensity mismatch as compared to the full-\wfqfmu absorption curve is observed if the J\&C dielectric function is employed. Remarkably, such a discrepancy increases as the shell thickness decreases. 

To quantitatively analyze surface-related near-field properties, $\rho^{\text{error}}$ values as a function of the permittivity function and $\Delta r$ are reported in \cref{table:prf-rho-dielectric}. Computed relative errors using BB and J\&C permittivities consistently increase by reducing the atomistic shell thickness, unlike those obtained by exploiting the Palik dielectric function, which remain almost constant (about 9\%). Notably, the best results (i.e. lowest errors) are obtained by employing BB, with the exception of the case $\Delta r = 7.93$ \AA~ for which Palik gives the best results.  J\&C data substantially deviate from reference \wfqfmu values. On the other hand, BB emerges as the most robust permittivity function model to exploit in \wfqfmubem applications. 

\begin{table}[htbp!]
    \centering
    \begin{tabular}{ccccc}
    \toprule
        $\Delta r$ (\AA) & Dielectric Function & PRF (eV) & $\rho^{\text{error}}$(\%) \\
    \midrule
        \multirow{3}{*}{13.98} & BB & 3.45 & 3.49 \\
                               & Palik & 3.48 & 8.21 \\
                               & J\&C & 3.48 & 22.21 \\ \cdashline{1-4}
        \multirow{3}{*}{11.82} & BB & 3.44 & 5.45 \\
                               & Palik & 3.48 & 9.26 \\
                               & J\&C & 3.48 & 28.30 \\ \cdashline{1-4}
        \multirow{3}{*}{9.95} & BB & 3.44 & 6.77 \\
                               & Palik & 3.48 & 10.70 \\
                               & J\&C & 3.48 & 31.82 \\ \cdashline{1-4}
        \multirow{3}{*}{7.93} & BB & 3.43 & 10.42 \\
                               & Palik & 3.48 & 9.22 \\
                               & J\&C & 3.48 & 44.45 \\\cdashline{1-4}
                               &       &      &       \\
        Full-\wfqfmu    & -- &    3.47      & --     \\ 
    \bottomrule
    \end{tabular}
    \caption{PRFs and $\rho^{\text{error}}$ computed for the studied spherical Ag NP ($R = 40$ \AA) at the full-\wfqfmu and \wfqfmubem ($d=5.76$ \AA) levels, as a function of \wfqfmubem geometrical parameter $\Delta r$ and BEM dielectric function [Brendel-Bormann (BB), Palik, and Johnson and Christy (JC)].}
    \label{table:prf-rho-dielectric}
\end{table}

As a final validation, we investigate the dependence of absorption spectra on the number of \textit{tesserae} used to mesh the BEM core. Specifically, we consider four different tessellations of the spherical BEM core (2990, 1948, 1018, and 390 \textit{tesserae}), while keeping the other model parameters fixed to the best set resulting from the previous analysis ($d = 5.76$ \AA, $\Delta r = 13.98$ \AA, BB permittivity function). Numerical results are reported in Fig. S4 and Table S2 of the \sm. Interestingly, the discretization of the core only minimally affects the plasmonic response (spectra, density-differences plots, and $\rho^{\text{error}}$), while yielding substantial computational \% speed-ups, up to 56.68\% for 390  \textit{tesserae}. To conclude, the best parameter set, which guarantees the best compromise between computational cost and accuracy, is defined by: $d = 5.76$ \AA, $\Delta r = 13.98$ \AA, BB permittivity function, 390 \emph{tesserae}.



\subsection{Molecular Plasmonics: QM/\texorpdfstring{\wfqfmubem}{} for SERS}\label{sec:qm/classical-sers}

\begin{figure*}[htb!]
        \centering
        \includegraphics[width=0.95\textwidth]{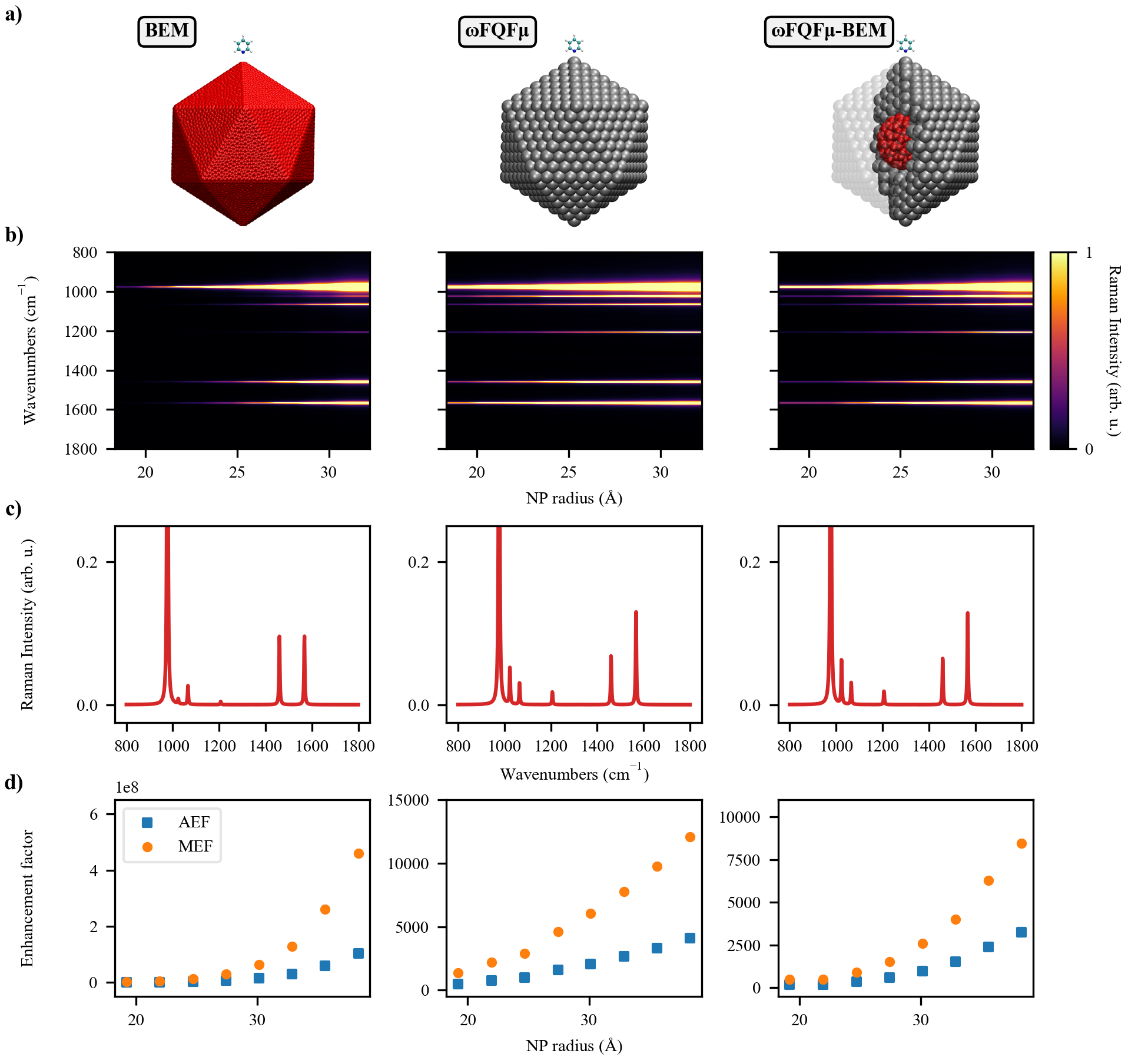}
        \caption{Graphical depiction of pyridine adsorbed on Ag Ih systems described at the full-BEM (left), full-\wfqfmu (center), and \wfqfmubem (right) levels. b) Normalized pyridine SERS spectra as a function of the NP radius. c) Normalized SERS of pyridine adsorbed on the largest NP. d) AEF and MEF as a function of the NP radius. BEM response is described with the BB dielectric function, and \wfqfmubem Ih systems are characterized by $\textit{d}=5.76$ \AA~.}
        \label{fig:sers-d5.76-bb}
\end{figure*}

As mentioned in the Introduction, and specified in Section \ref{theoqmm}, \wfqfmubem can be coupled to a QM Hamiltonian in a QM/MM fashion, and extended to compute SERS spectra. This section showcases the potentialities of the method to compute SERS spectra of pyridine near the tip of complex-shaped NPs. To this end, Ag icosahedral (Ih) NPs of increasing size (from 1.9 nm to 3.8 nm) are selected, and additional calculations modeling the molecule-NP system at the fully atomistic (QM/\wfqfmu) and fully implicit (QM/BEM) levels are discussed. A graphical representation highlighting the structural differences between the various methods is given in \cref{fig:sers-d5.76-bb}, panel a, whereas geometrical parameters are shown in Table S3 in the \sm. To fully characterize \wfqfmubem Ih systems and similarly to spherical NPs, four parameters are exploited (the BEM radius $r_{BEM}$, the BEM-\wfqfmu distance $d$, the average thickness of the \wfqfmu shell $\Delta r$ and the NP radius $R$), while the equivalent full-\wfqfmu and full-BEM structures are described by a single radius parameter ($R$). Detailed data on the geometrical features and PRFs of the systems are gathered in Tables S3 to S5 of the \sm.

In \cref{fig:sers-d5.76-bb}, the SERS signal of pyridine on Ag Ih NPs is studied as a function of the NP radius. Pyridine is adsorbed perpendicularly to the NP surface at a distance of 3 \AA, with the N atom lying closest to the NP. In agreement with the preliminary analysis in \cref{sec:dielectric-comparison}, the BB dielectric function is employed, and $d$ is set to $5.76$ \AA. \Cref{fig:sers-d5.76-bb}b illustrates SERS intensities as a function of the NP radius, normalized with respect to the largest signal exhibited by each model. In all cases, this corresponds to the SERS signal of pyridine interacting with the largest Ih NP. Notably, for all models only the signals associated with specific normal modes are enhanced (see Fig. S6 in the \sm for their graphical depiction). This is due to the fact that such vibrations are related to an orthogonal movement of pyridine atoms with respect to the NP surface. As a consequence, they experience the largest field gradient variation and, are therefore preferentially enhanced, as has recently been shown by some of us.\cite{lafiosca2023classical} The analysis of the results reveals significant similarities between QM/\wfqfmubem and QM/\wfqfmu approaches, while also highlighting substantial discrepancies by using QM/BEM. Specifically, the observed Raman peaks are consistent across all models when pyridine is adsorbed on the largest NPs (radius $>$ 2.7 nm). On the contrary, the smallest full-BEM structures (radius $<$ 2.5 nm) substantially deviate from the full atomistic picture (see \cref{fig:sers-d5.76-bb}b). To better compare the three approaches, \cref{fig:sers-d5.76-bb}c presents normalized Raman spectra for the largest NP studied using each theoretical framework. The spectra depicted in \cref{fig:sers-d5.76-bb}c show an almost perfect alignment of the relative Raman intensities between \wfqfmubem and full-\wfqfmu NPs. In contrast, substantial discrepancies are evident for full-BEM structures, particularly for the Raman peaks centered at 1024, 1207,1460 and 1568 cm$^{-1}$. 

To comprehensively analyze the performance of the various models, \cref{fig:sers-d5.76-bb}d collects AEF and MEF descriptors calculated as a function of the NP radius. For all methods, such indices are monotonically increasing. The reference QM/\wfqfmu approach predicts a maximum AEF and MEF of about 5000 and 12000, respectively, thus displaying an overall enhancement of about 10$^3$. QM/BEM shows a substantial quantitative mismatch, providing AEF and MEF that are completely overestimated by several orders of magnitude (about 10$^8$). Conversely, QM/\wfqfmubem presents a clear trend of quantitative similarity with the reference fully atomistic results, predicts AEF and MEF of the same order of magnitude and only deviating from the reference of a factor of about 1.35 for the largest structure.
%

It is worth remarking that the QM/BEM trends arising from \cref{fig:sers-d5.76-bb} can be substantially affected by varying the dielectric functions (J\&C and Palik). These results are reported in Sec. S2.2 of the \sm, and show that J\&C permittivity yields a complete modification of the spectral response for the largest NP (see Figs. S7 and S10 of the \sm), also affecting computed AEF and MEF values, further magnifying the quantitative discrepancies compared to the fully atomistic \wfqfmu description. Conversely, the QM/\wfqfmubem spectra remain consistent regardless of the chosen parameters, but are crucially affected by a variation in the BEM-\wfqfmu distance $d$. In fact, by setting $d=2.88$ \AA~, inconsistent trends in both AEF and MEF are reported as the system size increases (see Figs. S9 to S11 of the \sm). This behavior completely aligns with our findings from the study on $\rho^{\text{error}}$ of spherical \wfqfmubem NPs (see \cref{sec:thickness-dependence}). 

We conclude this section by remarking on the enhanced computational efficiency of QM/\wfqfmubem method as compared to the fully atomistic \wfqfmu description (see Table S6 of the \sm for the numerical \% speed-up). Notably, the computational saving is particularly evident for the largest systems (radius $>$ 3 nm), for which, as previously discussed, an improved quantitative and qualitative description of the SERS response is also obtained. These findings underscore the potential of \wfqfmubem for its application to large systems, where atomistic details at their surface play a major role in their plasmonic features, offering a substantial reduction in computational cost, up to 58\%.

\section{Conclusions}\label{sec:conclusions}

This work has presented a novel multiscale implicit-atomistic approach (\wfqfmubem) to investigate the plasmonic response of metal NPs. \wfqfmubem combines a spherical implicit continuum core embedded within an atomistically defined shell, preserving the essential plasmonic response at the NP surface. The method is theoretically constructed by integrating the fully atomistic \wfqfmu and implicit BEM methodologies for plasmonics and constitutes the first attempt to merge atomistic and continuum descriptions for the study of plasmonic substrates in the context of classical electrodynamics. 
To validate the \wfqfmubem model, we have initially demonstrated its capacity to reproduce the absorption cross-section of spherical Ag nanoparticles, taking the fully atomistic \wfqfmu method as a reference. Specifically, we have presented a comprehensive analysis of the different factors influencing \wfqfmubem response: thickness of the atomistic shell, BEM-\wfqfmu distance, BEM dielectric function, and discretization of the continuum core. Our in-depth analysis suggests that the best results at the implicit continuum-atomistic \wfqfmubem level for Ag NPs are obtained by setting the BEM-\wfqfmu distance to twice the nearest neighbor distance and using the Brendel-Bormann dielectric function while keeping the outer shells described atomistically. 

The coupling of \wfqfmubem to a QM region has been presented, giving rise to the QM/\wfqfmubem approach, which is able to describe molecular plasmonics, i.e. how plasmons affect the molecular electronic structure and the resulting spectral signals. To showcase the potentialities of the method, QM/\wfqfmubem has been challenged to reproduce the SERS response of pyridine adsorbed on the tip of Ag Ih NPs. The quality of the method has been assessed by an in-depth comparison with fully atomistic QM/\wfqfmu (reference) and fully implicit QM/BEM spectra. Our results show that, while QM/BEM outcomes strongly differ, QM/\wfqfmubem can reproduce QM/\wfqfmu SERS spectra, specifically for the largest NP sizes. More in detail, an almost perfect agreement of normalized Raman intensities for the largest system is observed, along with a similar qualitative and quantitative outcome for AEF and MEF descriptors with increasing NP radius. As a result of this work, two main points may be highlighted: (i) an implicit description of the NP structure yields substantial discrepancies in simulated SERS spectra as compared to an atomistic description and should be generally avoided; (ii) it is crucial to describe at a full atomistic level the NP surface, while the core can be safely treated by an implicit description. Finally, the enhanced computational efficiency observed for the largest QM/\wfqfmubem structures yields a computational saving of more than half compared to QM/\wfqfmu. Therefore, large, complex-shaped systems can be afforded. 
To end the presentation, it is worth mentioning that both BEM and \wfqfmu models are formulated in the quasistatic regime, i.e. retardation effects are discarded. Therefore, they appropriately treat systems of dimension significantly smaller than that of the incident wavelength. Also, in \wfqfmubem the \wfqfmu and BEM layers interact via electrostatic forces, i.e. charge transfer is inhibited. The above two limitations of the approach will deserve careful investigation in future communications.

\section*{Conflicts of interest}

There are no conflicts to declare.

\section{Acknowledgements}
The authors acknowledge Prof. Stefano Corni (University of Padova) for fruitful discussions. We gratefully acknowledge the Center for High-Performance Computing (CHPC) at SNS for providing the computational infrastructure. This work has received funding from the European Research Council (ERC) under the European Union’s Horizon 2020 research and innovation program (grant agreement N. 818064).

\section{Electronic Supplementary Material}

Details on \wfqfmubem equations. Geometrical parameters, PRFs, $\rho^{\text{error}}$, and computational speed-ups. Defined volume for $\rho^{\text{error}}$ calculations. Further results on classical and QM/classical responses as a function of core-to-shell distance, dielectric function, and BEM core discretization for spherical and Ih core-shell systems. Graphical representation of pyridine's normal modes.

\section{Data Availability}

The data that support the findings of this study are available from the corresponding authors upon reasonable request.

\bibliography{biblio}

\begin{thebibliography}{104}%
\makeatletter
\providecommand \@ifxundefined [1]{%
 \@ifx{#1\undefined}
}%
\providecommand \@ifnum [1]{%
 \ifnum #1\expandafter \@firstoftwo
 \else \expandafter \@secondoftwo
 \fi
}%
\providecommand \@ifx [1]{%
 \ifx #1\expandafter \@firstoftwo
 \else \expandafter \@secondoftwo
 \fi
}%
\providecommand \natexlab [1]{#1}%
\providecommand \enquote  [1]{``#1''}%
\providecommand \bibnamefont  [1]{#1}%
\providecommand \bibfnamefont [1]{#1}%
\providecommand \citenamefont [1]{#1}%
\providecommand \href@noop [0]{\@secondoftwo}%
\providecommand \href [0]{\begingroup \@sanitize@url \@href}%
\providecommand \@href[1]{\@@startlink{#1}\@@href}%
\providecommand \@@href[1]{\endgroup#1\@@endlink}%
\providecommand \@sanitize@url [0]{\catcode `\\12\catcode `\$12\catcode
  `\&12\catcode `\#12\catcode `\^12\catcode `\_12\catcode `\%12\relax}%
\providecommand \@@startlink[1]{}%
\providecommand \@@endlink[0]{}%
\providecommand \url  [0]{\begingroup\@sanitize@url \@url }%
\providecommand \@url [1]{\endgroup\@href {#1}{\urlprefix }}%
\providecommand \urlprefix  [0]{URL }%
\providecommand \Eprint [0]{\href }%
\providecommand \doibase [0]{http://dx.doi.org/}%
\providecommand \selectlanguage [0]{\@gobble}%
\providecommand \bibinfo  [0]{\@secondoftwo}%
\providecommand \bibfield  [0]{\@secondoftwo}%
\providecommand \translation [1]{[#1]}%
\providecommand \BibitemOpen [0]{}%
\providecommand \bibitemStop [0]{}%
\providecommand \bibitemNoStop [0]{.\EOS\space}%
\providecommand \EOS [0]{\spacefactor3000\relax}%
\providecommand \BibitemShut  [1]{\csname bibitem#1\endcsname}%
\let\auto@bib@innerbib\@empty
\bibitem [{\citenamefont {Giannini}\ \emph {et~al.}(2011)\citenamefont
  {Giannini}, \citenamefont {Fern{\'a}ndez-Dom{\'\i}nguez}, \citenamefont
  {Heck},\ and\ \citenamefont {Maier}}]{giannini2011plasmonic}%
  \BibitemOpen
  \bibfield  {author} {\bibinfo {author} {\bibfnamefont {V.}~\bibnamefont
  {Giannini}}, \bibinfo {author} {\bibfnamefont {A.~I.}\ \bibnamefont
  {Fern{\'a}ndez-Dom{\'\i}nguez}}, \bibinfo {author} {\bibfnamefont {S.~C.}\
  \bibnamefont {Heck}}, \ and\ \bibinfo {author} {\bibfnamefont {S.~A.}\
  \bibnamefont {Maier}},\ }\href@noop {} {\bibfield  {journal} {\bibinfo
  {journal} {Chem. Rev.}\ }\textbf {\bibinfo {volume} {111}},\ \bibinfo {pages}
  {3888} (\bibinfo {year} {2011})}\BibitemShut {NoStop}%
\bibitem [{\citenamefont {Maier}(2007)}]{maier2007plasmonics}%
  \BibitemOpen
  \bibfield  {author} {\bibinfo {author} {\bibfnamefont {S.~A.}\ \bibnamefont
  {Maier}},\ }\href@noop {} {\emph {\bibinfo {title} {Plasmonics: fundamentals
  and applications}}}\ (\bibinfo  {publisher} {Springer Science \& Business
  Media},\ \bibinfo {year} {2007})\BibitemShut {NoStop}%
\bibitem [{\citenamefont {Odom}\ and\ \citenamefont
  {Schatz}(2011)}]{odom2011introduction}%
  \BibitemOpen
  \bibfield  {author} {\bibinfo {author} {\bibfnamefont {T.~W.}\ \bibnamefont
  {Odom}}\ and\ \bibinfo {author} {\bibfnamefont {G.~C.}\ \bibnamefont
  {Schatz}},\ }\href@noop {} {\enquote {\bibinfo {title} {Introduction to
  plasmonics},}\ } (\bibinfo {year} {2011})\BibitemShut {NoStop}%
\bibitem [{\citenamefont {Kelly}\ \emph {et~al.}(2003)\citenamefont {Kelly},
  \citenamefont {Coronado}, \citenamefont {Zhao},\ and\ \citenamefont
  {Schatz}}]{kelly2003optical}%
  \BibitemOpen
  \bibfield  {author} {\bibinfo {author} {\bibfnamefont {K.~L.}\ \bibnamefont
  {Kelly}}, \bibinfo {author} {\bibfnamefont {E.}~\bibnamefont {Coronado}},
  \bibinfo {author} {\bibfnamefont {L.~L.}\ \bibnamefont {Zhao}}, \ and\
  \bibinfo {author} {\bibfnamefont {G.~C.}\ \bibnamefont {Schatz}},\
  }\href@noop {} {\enquote {\bibinfo {title} {The optical properties of metal
  nanoparticles: the influence of size, shape, and dielectric environment},}\ }
  (\bibinfo {year} {2003})\BibitemShut {NoStop}%
\bibitem [{\citenamefont {Coccia}\ \emph {et~al.}(2020)\citenamefont {Coccia},
  \citenamefont {Fregoni}, \citenamefont {Guido}, \citenamefont {Marsili},
  \citenamefont {Pipolo},\ and\ \citenamefont {Corni}}]{coccia2020hybrid}%
  \BibitemOpen
  \bibfield  {author} {\bibinfo {author} {\bibfnamefont {E.}~\bibnamefont
  {Coccia}}, \bibinfo {author} {\bibfnamefont {J.}~\bibnamefont {Fregoni}},
  \bibinfo {author} {\bibfnamefont {C.}~\bibnamefont {Guido}}, \bibinfo
  {author} {\bibfnamefont {M.}~\bibnamefont {Marsili}}, \bibinfo {author}
  {\bibfnamefont {S.}~\bibnamefont {Pipolo}}, \ and\ \bibinfo {author}
  {\bibfnamefont {S.}~\bibnamefont {Corni}},\ }\href@noop {} {\bibfield
  {journal} {\bibinfo  {journal} {J. Chem. Phys.}\ }\textbf {\bibinfo {volume}
  {153}},\ \bibinfo {pages} {200901} (\bibinfo {year} {2020})}\BibitemShut
  {NoStop}%
\bibitem [{\citenamefont {Willets}\ and\ \citenamefont
  {Van~Duyne}(2007)}]{willets2007localized}%
  \BibitemOpen
  \bibfield  {author} {\bibinfo {author} {\bibfnamefont {K.~A.}\ \bibnamefont
  {Willets}}\ and\ \bibinfo {author} {\bibfnamefont {R.~P.}\ \bibnamefont
  {Van~Duyne}},\ }\href@noop {} {\bibfield  {journal} {\bibinfo  {journal}
  {Annu. Rev. Phys. Chem.}\ }\textbf {\bibinfo {volume} {58}},\ \bibinfo
  {pages} {267} (\bibinfo {year} {2007})}\BibitemShut {NoStop}%
\bibitem [{\citenamefont {Zhang}\ \emph {et~al.}(2013)\citenamefont {Zhang},
  \citenamefont {Zhang}, \citenamefont {Dong}, \citenamefont {Jiang},
  \citenamefont {Zhang}, \citenamefont {Chen}, \citenamefont {Zhang},
  \citenamefont {Liao}, \citenamefont {Aizpurua}, \citenamefont {Luo} \emph
  {et~al.}}]{zhang2013chemical}%
  \BibitemOpen
  \bibfield  {author} {\bibinfo {author} {\bibfnamefont {R.}~\bibnamefont
  {Zhang}}, \bibinfo {author} {\bibfnamefont {Y.}~\bibnamefont {Zhang}},
  \bibinfo {author} {\bibfnamefont {Z.}~\bibnamefont {Dong}}, \bibinfo {author}
  {\bibfnamefont {S.}~\bibnamefont {Jiang}}, \bibinfo {author} {\bibfnamefont
  {C.}~\bibnamefont {Zhang}}, \bibinfo {author} {\bibfnamefont
  {L.}~\bibnamefont {Chen}}, \bibinfo {author} {\bibfnamefont {L.}~\bibnamefont
  {Zhang}}, \bibinfo {author} {\bibfnamefont {Y.}~\bibnamefont {Liao}},
  \bibinfo {author} {\bibfnamefont {J.}~\bibnamefont {Aizpurua}}, \bibinfo
  {author} {\bibfnamefont {Y.~e.}\ \bibnamefont {Luo}},  \emph {et~al.},\
  }\href@noop {} {\bibfield  {journal} {\bibinfo  {journal} {Nature}\ }\textbf
  {\bibinfo {volume} {498}},\ \bibinfo {pages} {82} (\bibinfo {year}
  {2013})}\BibitemShut {NoStop}%
\bibitem [{\citenamefont {Jiang}\ \emph {et~al.}(2015)\citenamefont {Jiang},
  \citenamefont {Zhang}, \citenamefont {Zhang}, \citenamefont {Hu},
  \citenamefont {Liao}, \citenamefont {Luo}, \citenamefont {Yang},
  \citenamefont {Dong},\ and\ \citenamefont {Hou}}]{jiang2015distinguishing}%
  \BibitemOpen
  \bibfield  {author} {\bibinfo {author} {\bibfnamefont {S.}~\bibnamefont
  {Jiang}}, \bibinfo {author} {\bibfnamefont {Y.}~\bibnamefont {Zhang}},
  \bibinfo {author} {\bibfnamefont {R.}~\bibnamefont {Zhang}}, \bibinfo
  {author} {\bibfnamefont {C.}~\bibnamefont {Hu}}, \bibinfo {author}
  {\bibfnamefont {M.}~\bibnamefont {Liao}}, \bibinfo {author} {\bibfnamefont
  {Y.}~\bibnamefont {Luo}}, \bibinfo {author} {\bibfnamefont {J.}~\bibnamefont
  {Yang}}, \bibinfo {author} {\bibfnamefont {Z.}~\bibnamefont {Dong}}, \ and\
  \bibinfo {author} {\bibfnamefont {J.}~\bibnamefont {Hou}},\ }\href@noop {}
  {\bibfield  {journal} {\bibinfo  {journal} {Nat. Nanotechnol.}\ }\textbf
  {\bibinfo {volume} {10}},\ \bibinfo {pages} {865} (\bibinfo {year}
  {2015})}\BibitemShut {NoStop}%
\bibitem [{\citenamefont {Chiang}\ \emph {et~al.}(2016)\citenamefont {Chiang},
  \citenamefont {Chen}, \citenamefont {Goubert}, \citenamefont {Chulhai},
  \citenamefont {Chen}, \citenamefont {Pozzi}, \citenamefont {Jiang},
  \citenamefont {Hersam}, \citenamefont {Seideman}, \citenamefont {Jensen},\
  and\ \citenamefont {Van~Duyne}}]{chiang2016conformational}%
  \BibitemOpen
  \bibfield  {author} {\bibinfo {author} {\bibfnamefont {N.}~\bibnamefont
  {Chiang}}, \bibinfo {author} {\bibfnamefont {X.}~\bibnamefont {Chen}},
  \bibinfo {author} {\bibfnamefont {G.}~\bibnamefont {Goubert}}, \bibinfo
  {author} {\bibfnamefont {D.~V.}\ \bibnamefont {Chulhai}}, \bibinfo {author}
  {\bibfnamefont {X.}~\bibnamefont {Chen}}, \bibinfo {author} {\bibfnamefont
  {E.~A.}\ \bibnamefont {Pozzi}}, \bibinfo {author} {\bibfnamefont
  {N.}~\bibnamefont {Jiang}}, \bibinfo {author} {\bibfnamefont {M.~C.}\
  \bibnamefont {Hersam}}, \bibinfo {author} {\bibfnamefont {T.}~\bibnamefont
  {Seideman}}, \bibinfo {author} {\bibfnamefont {L.}~\bibnamefont {Jensen}}, \
  and\ \bibinfo {author} {\bibfnamefont {R.~P.}\ \bibnamefont {Van~Duyne}},\
  }\href@noop {} {\bibfield  {journal} {\bibinfo  {journal} {Nano Lett.}\
  }\textbf {\bibinfo {volume} {16}},\ \bibinfo {pages} {7774} (\bibinfo {year}
  {2016})}\BibitemShut {NoStop}%
\bibitem [{\citenamefont {Langer}\ \emph {et~al.}(2019)\citenamefont {Langer},
  \citenamefont {Jimenez~de Aberasturi}, \citenamefont {Aizpurua},
  \citenamefont {Alvarez-Puebla}, \citenamefont {Augui{\'e}}, \citenamefont
  {Baumberg}, \citenamefont {Bazan}, \citenamefont {Bell}, \citenamefont
  {Boisen}, \citenamefont {Brolo} \emph {et~al.}}]{langer2019present}%
  \BibitemOpen
  \bibfield  {author} {\bibinfo {author} {\bibfnamefont {J.}~\bibnamefont
  {Langer}}, \bibinfo {author} {\bibfnamefont {D.}~\bibnamefont {Jimenez~de
  Aberasturi}}, \bibinfo {author} {\bibfnamefont {J.}~\bibnamefont {Aizpurua}},
  \bibinfo {author} {\bibfnamefont {R.~A.}\ \bibnamefont {Alvarez-Puebla}},
  \bibinfo {author} {\bibfnamefont {B.}~\bibnamefont {Augui{\'e}}}, \bibinfo
  {author} {\bibfnamefont {J.~J.}\ \bibnamefont {Baumberg}}, \bibinfo {author}
  {\bibfnamefont {G.~C.}\ \bibnamefont {Bazan}}, \bibinfo {author}
  {\bibfnamefont {S.~E.}\ \bibnamefont {Bell}}, \bibinfo {author}
  {\bibfnamefont {A.}~\bibnamefont {Boisen}}, \bibinfo {author} {\bibfnamefont
  {A.~G.}\ \bibnamefont {Brolo}},  \emph {et~al.},\ }\href@noop {} {\bibfield
  {journal} {\bibinfo  {journal} {ACS Nano}\ }\textbf {\bibinfo {volume}
  {14}},\ \bibinfo {pages} {28} (\bibinfo {year} {2019})}\BibitemShut {NoStop}%
\bibitem [{\citenamefont {Lafiosca}\ \emph {et~al.}(2023)\citenamefont
  {Lafiosca}, \citenamefont {Nicoli}, \citenamefont {Bonatti}, \citenamefont
  {Giovannini}, \citenamefont {Corni},\ and\ \citenamefont
  {Cappelli}}]{lafiosca2023classical}%
  \BibitemOpen
  \bibfield  {author} {\bibinfo {author} {\bibfnamefont {P.}~\bibnamefont
  {Lafiosca}}, \bibinfo {author} {\bibfnamefont {L.}~\bibnamefont {Nicoli}},
  \bibinfo {author} {\bibfnamefont {L.}~\bibnamefont {Bonatti}}, \bibinfo
  {author} {\bibfnamefont {T.}~\bibnamefont {Giovannini}}, \bibinfo {author}
  {\bibfnamefont {S.}~\bibnamefont {Corni}}, \ and\ \bibinfo {author}
  {\bibfnamefont {C.}~\bibnamefont {Cappelli}},\ }\href {\doibase
  10.1021/acs.jctc.3c00177} {\bibfield  {journal} {\bibinfo  {journal} {J.
  Chem. Theory Comput.}\ }\textbf {\bibinfo {volume} {19}},\ \bibinfo {pages}
  {3616} (\bibinfo {year} {2023})}\BibitemShut {NoStop}%
\bibitem [{\citenamefont {Zhao}, \citenamefont {Jensen},\ and\ \citenamefont
  {Schatz}(2006)}]{zhao2006pyridine}%
  \BibitemOpen
  \bibfield  {author} {\bibinfo {author} {\bibfnamefont {L.}~\bibnamefont
  {Zhao}}, \bibinfo {author} {\bibfnamefont {L.}~\bibnamefont {Jensen}}, \ and\
  \bibinfo {author} {\bibfnamefont {G.~C.}\ \bibnamefont {Schatz}},\
  }\href@noop {} {\bibfield  {journal} {\bibinfo  {journal} {J. Am. Chem.
  Soc.}\ }\textbf {\bibinfo {volume} {128}},\ \bibinfo {pages} {2911} (\bibinfo
  {year} {2006})}\BibitemShut {NoStop}%
\bibitem [{\citenamefont {Jensen}, \citenamefont {Aikens},\ and\ \citenamefont
  {Schatz}(2008)}]{jensen2008electronic}%
  \BibitemOpen
  \bibfield  {author} {\bibinfo {author} {\bibfnamefont {L.}~\bibnamefont
  {Jensen}}, \bibinfo {author} {\bibfnamefont {C.~M.}\ \bibnamefont {Aikens}},
  \ and\ \bibinfo {author} {\bibfnamefont {G.~C.}\ \bibnamefont {Schatz}},\
  }\href@noop {} {\bibfield  {journal} {\bibinfo  {journal} {Chem. Soc. Rev.}\
  }\textbf {\bibinfo {volume} {37}},\ \bibinfo {pages} {1061} (\bibinfo {year}
  {2008})}\BibitemShut {NoStop}%
\bibitem [{\citenamefont {Morton}\ and\ \citenamefont
  {Jensen}(2009)}]{morton2009understanding}%
  \BibitemOpen
  \bibfield  {author} {\bibinfo {author} {\bibfnamefont {S.~M.}\ \bibnamefont
  {Morton}}\ and\ \bibinfo {author} {\bibfnamefont {L.}~\bibnamefont
  {Jensen}},\ }\href@noop {} {\bibfield  {journal} {\bibinfo  {journal} {J. Am.
  Chem. Soc.}\ }\textbf {\bibinfo {volume} {131}},\ \bibinfo {pages} {4090}
  (\bibinfo {year} {2009})}\BibitemShut {NoStop}%
\bibitem [{\citenamefont {Morton}, \citenamefont {Silverstein},\ and\
  \citenamefont {Jensen}(2011)}]{morton2011theoretical}%
  \BibitemOpen
  \bibfield  {author} {\bibinfo {author} {\bibfnamefont {S.~M.}\ \bibnamefont
  {Morton}}, \bibinfo {author} {\bibfnamefont {D.~W.}\ \bibnamefont
  {Silverstein}}, \ and\ \bibinfo {author} {\bibfnamefont {L.}~\bibnamefont
  {Jensen}},\ }\href@noop {} {\bibfield  {journal} {\bibinfo  {journal} {Chem.
  Rev.}\ }\textbf {\bibinfo {volume} {111}},\ \bibinfo {pages} {3962} (\bibinfo
  {year} {2011})}\BibitemShut {NoStop}%
\bibitem [{\citenamefont {Sanchez-Gonzalez}, \citenamefont {Corni},\ and\
  \citenamefont {Mennucci}(2011)}]{gonzalez2011surface}%
  \BibitemOpen
  \bibfield  {author} {\bibinfo {author} {\bibfnamefont {A.}~\bibnamefont
  {Sanchez-Gonzalez}}, \bibinfo {author} {\bibfnamefont {S.}~\bibnamefont
  {Corni}}, \ and\ \bibinfo {author} {\bibfnamefont {B.}~\bibnamefont
  {Mennucci}},\ }\href {\doibase 10.1021/jp111196f} {\bibfield  {journal}
  {\bibinfo  {journal} {J. Phys. Chem. C}\ }\textbf {\bibinfo {volume} {115}},\
  \bibinfo {pages} {5450} (\bibinfo {year} {2011})}\BibitemShut {NoStop}%
\bibitem [{\citenamefont {Morton}\ and\ \citenamefont
  {Jensen}(2010)}]{morton2010discrete}%
  \BibitemOpen
  \bibfield  {author} {\bibinfo {author} {\bibfnamefont {S.~M.}\ \bibnamefont
  {Morton}}\ and\ \bibinfo {author} {\bibfnamefont {L.}~\bibnamefont
  {Jensen}},\ }\href@noop {} {\bibfield  {journal} {\bibinfo  {journal} {J.
  Chem. Phys.}\ }\textbf {\bibinfo {volume} {133}},\ \bibinfo {pages} {074103}
  (\bibinfo {year} {2010})}\BibitemShut {NoStop}%
\bibitem [{\citenamefont {Morton}\ and\ \citenamefont
  {Jensen}(2011)}]{morton2011discrete}%
  \BibitemOpen
  \bibfield  {author} {\bibinfo {author} {\bibfnamefont {S.~M.}\ \bibnamefont
  {Morton}}\ and\ \bibinfo {author} {\bibfnamefont {L.}~\bibnamefont
  {Jensen}},\ }\href@noop {} {\bibfield  {journal} {\bibinfo  {journal} {J.
  Chem. Phys.}\ }\textbf {\bibinfo {volume} {135}},\ \bibinfo {pages} {134103}
  (\bibinfo {year} {2011})}\BibitemShut {NoStop}%
\bibitem [{\citenamefont {Mie}(1908)}]{mie1908beitrage}%
  \BibitemOpen
  \bibfield  {author} {\bibinfo {author} {\bibfnamefont {G.}~\bibnamefont
  {Mie}},\ }\href@noop {} {\bibfield  {journal} {\bibinfo  {journal} {Ann.
  Phys.-Berlin}\ }\textbf {\bibinfo {volume} {330}},\ \bibinfo {pages} {377}
  (\bibinfo {year} {1908})}\BibitemShut {NoStop}%
\bibitem [{\citenamefont {Taflove}, \citenamefont {Hagness},\ and\
  \citenamefont {Piket-May}(2005)}]{taflove2005computational}%
  \BibitemOpen
  \bibfield  {author} {\bibinfo {author} {\bibfnamefont {A.}~\bibnamefont
  {Taflove}}, \bibinfo {author} {\bibfnamefont {S.~C.}\ \bibnamefont
  {Hagness}}, \ and\ \bibinfo {author} {\bibfnamefont {M.}~\bibnamefont
  {Piket-May}},\ }\href@noop {} {\emph {\bibinfo {title} {Computational
  electromagnetics: the finite-difference time-domain method}}}\ (\bibinfo
  {publisher} {Elsevier Amsterdam, The Netherlands},\ \bibinfo {year}
  {2005})\BibitemShut {NoStop}%
\bibitem [{\citenamefont {Garc{\'\i}a~de Abajo}\ and\ \citenamefont
  {Howie}(2002)}]{de2002retarded}%
  \BibitemOpen
  \bibfield  {author} {\bibinfo {author} {\bibfnamefont {F.~J.}\ \bibnamefont
  {Garc{\'\i}a~de Abajo}}\ and\ \bibinfo {author} {\bibfnamefont
  {A.}~\bibnamefont {Howie}},\ }\href@noop {} {\bibfield  {journal} {\bibinfo
  {journal} {Phys. Rev. B}\ }\textbf {\bibinfo {volume} {65}},\ \bibinfo
  {pages} {115418} (\bibinfo {year} {2002})}\BibitemShut {NoStop}%
\bibitem [{\citenamefont {Myroshnychenko}\ \emph {et~al.}(2008)\citenamefont
  {Myroshnychenko}, \citenamefont {Carb{\'o}-Argibay}, \citenamefont
  {Pastoriza-Santos}, \citenamefont {P{\'e}rez-Juste}, \citenamefont
  {Liz-Marz{\'a}n},\ and\ \citenamefont {Garc{\'\i}a~de
  Abajo}}]{myroshnychenko2008modeling}%
  \BibitemOpen
  \bibfield  {author} {\bibinfo {author} {\bibfnamefont {V.}~\bibnamefont
  {Myroshnychenko}}, \bibinfo {author} {\bibfnamefont {E.}~\bibnamefont
  {Carb{\'o}-Argibay}}, \bibinfo {author} {\bibfnamefont {I.}~\bibnamefont
  {Pastoriza-Santos}}, \bibinfo {author} {\bibfnamefont {J.}~\bibnamefont
  {P{\'e}rez-Juste}}, \bibinfo {author} {\bibfnamefont {L.~M.}\ \bibnamefont
  {Liz-Marz{\'a}n}}, \ and\ \bibinfo {author} {\bibfnamefont {F.~J.}\
  \bibnamefont {Garc{\'\i}a~de Abajo}},\ }\href@noop {} {\bibfield  {journal}
  {\bibinfo  {journal} {Adv. Mater.}\ }\textbf {\bibinfo {volume} {20}},\
  \bibinfo {pages} {4288} (\bibinfo {year} {2008})}\BibitemShut {NoStop}%
\bibitem [{\citenamefont {Mennucci}\ and\ \citenamefont
  {Corni}(2019)}]{mennucci2019multiscale}%
  \BibitemOpen
  \bibfield  {author} {\bibinfo {author} {\bibfnamefont {B.}~\bibnamefont
  {Mennucci}}\ and\ \bibinfo {author} {\bibfnamefont {S.}~\bibnamefont
  {Corni}},\ }\href@noop {} {\bibfield  {journal} {\bibinfo  {journal} {Nat.
  Rev. Chem.}\ }\textbf {\bibinfo {volume} {3}},\ \bibinfo {pages} {315}
  (\bibinfo {year} {2019})}\BibitemShut {NoStop}%
\bibitem [{\citenamefont {Bonatti}\ \emph {et~al.}(2020)\citenamefont
  {Bonatti}, \citenamefont {Gil}, \citenamefont {Giovannini}, \citenamefont
  {Corni},\ and\ \citenamefont {Cappelli}}]{bonatti2020plasmonic}%
  \BibitemOpen
  \bibfield  {author} {\bibinfo {author} {\bibfnamefont {L.}~\bibnamefont
  {Bonatti}}, \bibinfo {author} {\bibfnamefont {G.}~\bibnamefont {Gil}},
  \bibinfo {author} {\bibfnamefont {T.}~\bibnamefont {Giovannini}}, \bibinfo
  {author} {\bibfnamefont {S.}~\bibnamefont {Corni}}, \ and\ \bibinfo {author}
  {\bibfnamefont {C.}~\bibnamefont {Cappelli}},\ }\href@noop {} {\bibfield
  {journal} {\bibinfo  {journal} {Front. Chem.}\ }\textbf {\bibinfo {volume}
  {8}},\ \bibinfo {pages} {340} (\bibinfo {year} {2020})}\BibitemShut {NoStop}%
\bibitem [{\citenamefont {Jensen}\ and\ \citenamefont
  {Jensen}(2008)}]{jensen2008electrostatic}%
  \BibitemOpen
  \bibfield  {author} {\bibinfo {author} {\bibfnamefont {L.~L.}\ \bibnamefont
  {Jensen}}\ and\ \bibinfo {author} {\bibfnamefont {L.}~\bibnamefont
  {Jensen}},\ }\href {\doibase 10.1021/jp804116z} {\bibfield  {journal}
  {\bibinfo  {journal} {J. Phys. Chem. C}\ }\textbf {\bibinfo {volume} {112}},\
  \bibinfo {pages} {15697} (\bibinfo {year} {2008})}\BibitemShut {NoStop}%
\bibitem [{\citenamefont {Jensen}\ and\ \citenamefont
  {Jensen}(2009)}]{jensen2009atomistic}%
  \BibitemOpen
  \bibfield  {author} {\bibinfo {author} {\bibfnamefont {L.~L.}\ \bibnamefont
  {Jensen}}\ and\ \bibinfo {author} {\bibfnamefont {L.}~\bibnamefont
  {Jensen}},\ }\href@noop {} {\bibfield  {journal} {\bibinfo  {journal} {J.
  Phys. Chem. C}\ }\textbf {\bibinfo {volume} {113}},\ \bibinfo {pages} {15182}
  (\bibinfo {year} {2009})}\BibitemShut {NoStop}%
\bibitem [{\citenamefont {Zakomirnyi}\ \emph {et~al.}(2019)\citenamefont
  {Zakomirnyi}, \citenamefont {Rinkevicius}, \citenamefont {Baryshnikov},
  \citenamefont {S{\o}rensen},\ and\ \citenamefont
  {{\AA}gren}}]{zakomirnyi2019extended}%
  \BibitemOpen
  \bibfield  {author} {\bibinfo {author} {\bibfnamefont {V.~I.}\ \bibnamefont
  {Zakomirnyi}}, \bibinfo {author} {\bibfnamefont {Z.}~\bibnamefont
  {Rinkevicius}}, \bibinfo {author} {\bibfnamefont {G.~V.}\ \bibnamefont
  {Baryshnikov}}, \bibinfo {author} {\bibfnamefont {L.~K.}\ \bibnamefont
  {S{\o}rensen}}, \ and\ \bibinfo {author} {\bibfnamefont {H.}~\bibnamefont
  {{\AA}gren}},\ }\href@noop {} {\bibfield  {journal} {\bibinfo  {journal} {J.
  Phys. Chem. C}\ }\textbf {\bibinfo {volume} {123}},\ \bibinfo {pages} {28867}
  (\bibinfo {year} {2019})}\BibitemShut {NoStop}%
\bibitem [{\citenamefont {Zakomirnyi}\ \emph {et~al.}(2020)\citenamefont
  {Zakomirnyi}, \citenamefont {Rasskazov}, \citenamefont {S{\o}rensen},
  \citenamefont {Carney}, \citenamefont {Rinkevicius},\ and\ \citenamefont
  {{\AA}gren}}]{zakomirnyi2020plasmonic}%
  \BibitemOpen
  \bibfield  {author} {\bibinfo {author} {\bibfnamefont {V.~I.}\ \bibnamefont
  {Zakomirnyi}}, \bibinfo {author} {\bibfnamefont {I.~L.}\ \bibnamefont
  {Rasskazov}}, \bibinfo {author} {\bibfnamefont {L.~K.}\ \bibnamefont
  {S{\o}rensen}}, \bibinfo {author} {\bibfnamefont {P.~S.}\ \bibnamefont
  {Carney}}, \bibinfo {author} {\bibfnamefont {Z.}~\bibnamefont {Rinkevicius}},
  \ and\ \bibinfo {author} {\bibfnamefont {H.}~\bibnamefont {{\AA}gren}},\
  }\href@noop {} {\bibfield  {journal} {\bibinfo  {journal} {Phys. Chem. Chem.
  Phys.}\ }\textbf {\bibinfo {volume} {22}},\ \bibinfo {pages} {13467}
  (\bibinfo {year} {2020})}\BibitemShut {NoStop}%
\bibitem [{\citenamefont {Giovannini}\ \emph
  {et~al.}(2019{\natexlab{a}})\citenamefont {Giovannini}, \citenamefont {Rosa},
  \citenamefont {Corni},\ and\ \citenamefont
  {Cappelli}}]{giovannini2019classical}%
  \BibitemOpen
  \bibfield  {author} {\bibinfo {author} {\bibfnamefont {T.}~\bibnamefont
  {Giovannini}}, \bibinfo {author} {\bibfnamefont {M.}~\bibnamefont {Rosa}},
  \bibinfo {author} {\bibfnamefont {S.}~\bibnamefont {Corni}}, \ and\ \bibinfo
  {author} {\bibfnamefont {C.}~\bibnamefont {Cappelli}},\ }\href@noop {}
  {\bibfield  {journal} {\bibinfo  {journal} {Nanoscale}\ }\textbf {\bibinfo
  {volume} {11}},\ \bibinfo {pages} {6004} (\bibinfo {year}
  {2019}{\natexlab{a}})}\BibitemShut {NoStop}%
\bibitem [{\citenamefont {Giovannini}\ \emph {et~al.}(2020)\citenamefont
  {Giovannini}, \citenamefont {Bonatti}, \citenamefont {Polini},\ and\
  \citenamefont {Cappelli}}]{giovannini2020graphene}%
  \BibitemOpen
  \bibfield  {author} {\bibinfo {author} {\bibfnamefont {T.}~\bibnamefont
  {Giovannini}}, \bibinfo {author} {\bibfnamefont {L.}~\bibnamefont {Bonatti}},
  \bibinfo {author} {\bibfnamefont {M.}~\bibnamefont {Polini}}, \ and\ \bibinfo
  {author} {\bibfnamefont {C.}~\bibnamefont {Cappelli}},\ }\href@noop {}
  {\bibfield  {journal} {\bibinfo  {journal} {J. Phys. Chem. Lett.}\ }\textbf
  {\bibinfo {volume} {11}},\ \bibinfo {pages} {7595} (\bibinfo {year}
  {2020})}\BibitemShut {NoStop}%
\bibitem [{\citenamefont {Lafiosca}\ \emph {et~al.}(2021)\citenamefont
  {Lafiosca}, \citenamefont {Giovannini}, \citenamefont {Benzi},\ and\
  \citenamefont {Cappelli}}]{lafiosca2021going}%
  \BibitemOpen
  \bibfield  {author} {\bibinfo {author} {\bibfnamefont {P.}~\bibnamefont
  {Lafiosca}}, \bibinfo {author} {\bibfnamefont {T.}~\bibnamefont
  {Giovannini}}, \bibinfo {author} {\bibfnamefont {M.}~\bibnamefont {Benzi}}, \
  and\ \bibinfo {author} {\bibfnamefont {C.}~\bibnamefont {Cappelli}},\ }\href
  {\doibase 10.1021/acs.jpcc.1c04716} {\bibfield  {journal} {\bibinfo
  {journal} {J. Phys. Chem. C}\ }\textbf {\bibinfo {volume} {125}},\ \bibinfo
  {pages} {23848} (\bibinfo {year} {2021})}\BibitemShut {NoStop}%
\bibitem [{\citenamefont {Bonatti}\ \emph {et~al.}(2022)\citenamefont
  {Bonatti}, \citenamefont {Nicoli}, \citenamefont {Giovannini},\ and\
  \citenamefont {Cappelli}}]{bonatti2022silico}%
  \BibitemOpen
  \bibfield  {author} {\bibinfo {author} {\bibfnamefont {L.}~\bibnamefont
  {Bonatti}}, \bibinfo {author} {\bibfnamefont {L.}~\bibnamefont {Nicoli}},
  \bibinfo {author} {\bibfnamefont {T.}~\bibnamefont {Giovannini}}, \ and\
  \bibinfo {author} {\bibfnamefont {C.}~\bibnamefont {Cappelli}},\ }\href@noop
  {} {\bibfield  {journal} {\bibinfo  {journal} {Nanoscale Adv.}\ }\textbf
  {\bibinfo {volume} {4}},\ \bibinfo {pages} {2294} (\bibinfo {year}
  {2022})}\BibitemShut {NoStop}%
\bibitem [{\citenamefont {Giovannini}\ \emph {et~al.}(2022)\citenamefont
  {Giovannini}, \citenamefont {Bonatti}, \citenamefont {Lafiosca},
  \citenamefont {Nicoli}, \citenamefont {Castagnola}, \citenamefont {Illobre},
  \citenamefont {Corni},\ and\ \citenamefont {Cappelli}}]{giovannini2022we}%
  \BibitemOpen
  \bibfield  {author} {\bibinfo {author} {\bibfnamefont {T.}~\bibnamefont
  {Giovannini}}, \bibinfo {author} {\bibfnamefont {L.}~\bibnamefont {Bonatti}},
  \bibinfo {author} {\bibfnamefont {P.}~\bibnamefont {Lafiosca}}, \bibinfo
  {author} {\bibfnamefont {L.}~\bibnamefont {Nicoli}}, \bibinfo {author}
  {\bibfnamefont {M.}~\bibnamefont {Castagnola}}, \bibinfo {author}
  {\bibfnamefont {P.~G.}\ \bibnamefont {Illobre}}, \bibinfo {author}
  {\bibfnamefont {S.}~\bibnamefont {Corni}}, \ and\ \bibinfo {author}
  {\bibfnamefont {C.}~\bibnamefont {Cappelli}},\ }\href {\doibase
  10.1021/acsphotonics.2c00761} {\bibfield  {journal} {\bibinfo  {journal} {ACS
  Photonics}\ }\textbf {\bibinfo {volume} {9}},\ \bibinfo {pages} {3025}
  (\bibinfo {year} {2022})}\BibitemShut {NoStop}%
\bibitem [{\citenamefont {Zanotto}\ \emph {et~al.}(2023)\citenamefont
  {Zanotto}, \citenamefont {Bonatti}, \citenamefont {Pantano}, \citenamefont
  {Mi\u{s}eikis}, \citenamefont {Speranza}, \citenamefont {Giovannini},
  \citenamefont {Coletti}, \citenamefont {Cappelli}, \citenamefont
  {Tredicucci},\ and\ \citenamefont {Toncelli}}]{zanotto2023strain}%
  \BibitemOpen
  \bibfield  {author} {\bibinfo {author} {\bibfnamefont {S.}~\bibnamefont
  {Zanotto}}, \bibinfo {author} {\bibfnamefont {L.}~\bibnamefont {Bonatti}},
  \bibinfo {author} {\bibfnamefont {M.~F.}\ \bibnamefont {Pantano}}, \bibinfo
  {author} {\bibfnamefont {V.}~\bibnamefont {Mi\u{s}eikis}}, \bibinfo {author}
  {\bibfnamefont {G.}~\bibnamefont {Speranza}}, \bibinfo {author}
  {\bibfnamefont {T.}~\bibnamefont {Giovannini}}, \bibinfo {author}
  {\bibfnamefont {C.}~\bibnamefont {Coletti}}, \bibinfo {author} {\bibfnamefont
  {C.}~\bibnamefont {Cappelli}}, \bibinfo {author} {\bibfnamefont
  {A.}~\bibnamefont {Tredicucci}}, \ and\ \bibinfo {author} {\bibfnamefont
  {A.}~\bibnamefont {Toncelli}},\ }\href@noop {} {\bibfield  {journal}
  {\bibinfo  {journal} {ACS Photonics}\ }\textbf {\bibinfo {volume} {10}},\
  \bibinfo {pages} {394} (\bibinfo {year} {2023})}\BibitemShut {NoStop}%
\bibitem [{\citenamefont {Nicoli}\ \emph {et~al.}(2023)\citenamefont {Nicoli},
  \citenamefont {Lafiosca}, \citenamefont {Grobas~Illobre}, \citenamefont
  {Bonatti}, \citenamefont {Giovannini},\ and\ \citenamefont
  {Cappelli}}]{nicoli2023fully}%
  \BibitemOpen
  \bibfield  {author} {\bibinfo {author} {\bibfnamefont {L.}~\bibnamefont
  {Nicoli}}, \bibinfo {author} {\bibfnamefont {P.}~\bibnamefont {Lafiosca}},
  \bibinfo {author} {\bibfnamefont {P.}~\bibnamefont {Grobas~Illobre}},
  \bibinfo {author} {\bibfnamefont {L.}~\bibnamefont {Bonatti}}, \bibinfo
  {author} {\bibfnamefont {T.}~\bibnamefont {Giovannini}}, \ and\ \bibinfo
  {author} {\bibfnamefont {C.}~\bibnamefont {Cappelli}},\ }\href {\doibase
  10.3389/fphot.2023.1199598} {\bibfield  {journal} {\bibinfo  {journal}
  {Front. Photon.}\ }\textbf {\bibinfo {volume} {4}},\ \bibinfo {pages} {2673}
  (\bibinfo {year} {2023})}\BibitemShut {NoStop}%
\bibitem [{\citenamefont {Lafiosca}\ \emph {et~al.}(2024)\citenamefont
  {Lafiosca}, \citenamefont {Nicoli}, \citenamefont {Pipolo}, \citenamefont
  {Corni}, \citenamefont {Giovannini},\ and\ \citenamefont
  {Cappelli}}]{lafiosca2024real}%
  \BibitemOpen
  \bibfield  {author} {\bibinfo {author} {\bibfnamefont {P.}~\bibnamefont
  {Lafiosca}}, \bibinfo {author} {\bibfnamefont {L.}~\bibnamefont {Nicoli}},
  \bibinfo {author} {\bibfnamefont {S.}~\bibnamefont {Pipolo}}, \bibinfo
  {author} {\bibfnamefont {S.}~\bibnamefont {Corni}}, \bibinfo {author}
  {\bibfnamefont {T.}~\bibnamefont {Giovannini}}, \ and\ \bibinfo {author}
  {\bibfnamefont {C.}~\bibnamefont {Cappelli}},\ }\href@noop {} {\bibfield
  {journal} {\bibinfo  {journal} {J. Phys. Chem. C}\ }\textbf {\bibinfo
  {volume} {128}},\ \bibinfo {pages} {17513} (\bibinfo {year}
  {2024})}\BibitemShut {NoStop}%
\bibitem [{\citenamefont {Nicoli}\ \emph {et~al.}(2024)\citenamefont {Nicoli},
  \citenamefont {Sodomaco}, \citenamefont {Lafiosca}, \citenamefont
  {Giovannini},\ and\ \citenamefont {Cappelli}}]{nicoli2024atomistic}%
  \BibitemOpen
  \bibfield  {author} {\bibinfo {author} {\bibfnamefont {L.}~\bibnamefont
  {Nicoli}}, \bibinfo {author} {\bibfnamefont {S.}~\bibnamefont {Sodomaco}},
  \bibinfo {author} {\bibfnamefont {P.}~\bibnamefont {Lafiosca}}, \bibinfo
  {author} {\bibfnamefont {T.}~\bibnamefont {Giovannini}}, \ and\ \bibinfo
  {author} {\bibfnamefont {C.}~\bibnamefont {Cappelli}},\ }\href@noop {}
  {\bibfield  {journal} {\bibinfo  {journal} {ACS Phys. Chem. Au}\ ,\ \bibinfo
  {pages} {DOI: 10.1021/acsphyschemau.4c00052}} (\bibinfo {year}
  {2024})}\BibitemShut {NoStop}%
\bibitem [{\citenamefont {Corni}\ and\ \citenamefont
  {Tomasi}(2001{\natexlab{a}})}]{corni2001enhanced}%
  \BibitemOpen
  \bibfield  {author} {\bibinfo {author} {\bibfnamefont {S.}~\bibnamefont
  {Corni}}\ and\ \bibinfo {author} {\bibfnamefont {J.}~\bibnamefont {Tomasi}},\
  }\href@noop {} {\bibfield  {journal} {\bibinfo  {journal} {J. Chem. Phys.}\
  }\textbf {\bibinfo {volume} {114}},\ \bibinfo {pages} {3739} (\bibinfo {year}
  {2001}{\natexlab{a}})}\BibitemShut {NoStop}%
\bibitem [{\citenamefont {Corni}\ and\ \citenamefont
  {Tomasi}(2001{\natexlab{b}})}]{corni2001theoretical}%
  \BibitemOpen
  \bibfield  {author} {\bibinfo {author} {\bibfnamefont {S.}~\bibnamefont
  {Corni}}\ and\ \bibinfo {author} {\bibfnamefont {J.}~\bibnamefont {Tomasi}},\
  }\href@noop {} {\bibfield  {journal} {\bibinfo  {journal} {Chem. Phys.
  Lett.}\ }\textbf {\bibinfo {volume} {342}},\ \bibinfo {pages} {135} (\bibinfo
  {year} {2001}{\natexlab{b}})}\BibitemShut {NoStop}%
\bibitem [{\citenamefont {Corni}\ and\ \citenamefont
  {Tomasi}(2002{\natexlab{a}})}]{corni2002surface}%
  \BibitemOpen
  \bibfield  {author} {\bibinfo {author} {\bibfnamefont {S.}~\bibnamefont
  {Corni}}\ and\ \bibinfo {author} {\bibfnamefont {J.}~\bibnamefont {Tomasi}},\
  }\href@noop {} {\bibfield  {journal} {\bibinfo  {journal} {J. Chem. Phys.}\
  }\textbf {\bibinfo {volume} {116}},\ \bibinfo {pages} {1156} (\bibinfo {year}
  {2002}{\natexlab{a}})}\BibitemShut {NoStop}%
\bibitem [{\citenamefont {Corni}\ and\ \citenamefont
  {Tomasi}(2002{\natexlab{b}})}]{corni2002erratum}%
  \BibitemOpen
  \bibfield  {author} {\bibinfo {author} {\bibfnamefont {S.}~\bibnamefont
  {Corni}}\ and\ \bibinfo {author} {\bibfnamefont {J.}~\bibnamefont {Tomasi}},\
  }\href@noop {} {\bibfield  {journal} {\bibinfo  {journal} {Chem. Phys.
  Lett.}\ }\textbf {\bibinfo {volume} {365}},\ \bibinfo {pages} {552} (\bibinfo
  {year} {2002}{\natexlab{b}})}\BibitemShut {NoStop}%
\bibitem [{\citenamefont {Corni}\ and\ \citenamefont
  {Tomasi}(2006)}]{corni2006studying}%
  \BibitemOpen
  \bibfield  {author} {\bibinfo {author} {\bibfnamefont {S.}~\bibnamefont
  {Corni}}\ and\ \bibinfo {author} {\bibfnamefont {J.}~\bibnamefont {Tomasi}},\
  }in\ \href@noop {} {\emph {\bibinfo {booktitle} {Surface-Enhanced Raman
  Scattering}}}\ (\bibinfo  {publisher} {Springer},\ \bibinfo {year} {2006})\
  pp.\ \bibinfo {pages} {105--123}\BibitemShut {NoStop}%
\bibitem [{\citenamefont {Romanelli}, \citenamefont {Dall’Osto},\ and\
  \citenamefont {Corni}(2021)}]{corni2021role}%
  \BibitemOpen
  \bibfield  {author} {\bibinfo {author} {\bibfnamefont {M.}~\bibnamefont
  {Romanelli}}, \bibinfo {author} {\bibfnamefont {G.}~\bibnamefont
  {Dall’Osto}}, \ and\ \bibinfo {author} {\bibfnamefont {S.}~\bibnamefont
  {Corni}},\ }\href {\doibase 10.1063/5.0066758} {\bibfield  {journal}
  {\bibinfo  {journal} {J. Chem. Phys.}\ }\textbf {\bibinfo {volume} {155}},\
  \bibinfo {pages} {214304} (\bibinfo {year} {2021})}\BibitemShut {NoStop}%
\bibitem [{\citenamefont {Coane}\ \emph {et~al.}(2024)\citenamefont {Coane},
  \citenamefont {Romanelli}, \citenamefont {Dall'Osto}, \citenamefont
  {Di~Felice},\ and\ \citenamefont {Corni}}]{corni2024unraveling}%
  \BibitemOpen
  \bibfield  {author} {\bibinfo {author} {\bibfnamefont {C.~V.}\ \bibnamefont
  {Coane}}, \bibinfo {author} {\bibfnamefont {M.}~\bibnamefont {Romanelli}},
  \bibinfo {author} {\bibfnamefont {G.}~\bibnamefont {Dall'Osto}}, \bibinfo
  {author} {\bibfnamefont {R.}~\bibnamefont {Di~Felice}}, \ and\ \bibinfo
  {author} {\bibfnamefont {S.}~\bibnamefont {Corni}},\ }\href {\doibase
  10.1038/s42004-024-01118-1} {\bibfield  {journal} {\bibinfo  {journal}
  {Commun. Chem.}\ }\textbf {\bibinfo {volume} {7}},\ \bibinfo {pages} {32}
  (\bibinfo {year} {2024})}\BibitemShut {NoStop}%
\bibitem [{\citenamefont {Hohenester}\ and\ \citenamefont
  {Trügler}(2012)}]{trugler2012}%
  \BibitemOpen
  \bibfield  {author} {\bibinfo {author} {\bibfnamefont {U.}~\bibnamefont
  {Hohenester}}\ and\ \bibinfo {author} {\bibfnamefont {A.}~\bibnamefont
  {Trügler}},\ }\href {\doibase https://doi.org/10.1016/j.cpc.2011.09.009}
  {\bibfield  {journal} {\bibinfo  {journal} {Comput. Phys. Commun.}\ }\textbf
  {\bibinfo {volume} {183}},\ \bibinfo {pages} {370} (\bibinfo {year}
  {2012})}\BibitemShut {NoStop}%
\bibitem [{\citenamefont {Casida}(1995)}]{casida1995time}%
  \BibitemOpen
  \bibfield  {author} {\bibinfo {author} {\bibfnamefont {M.~E.}\ \bibnamefont
  {Casida}},\ }in\ \href@noop {} {\emph {\bibinfo {booktitle} {Recent Advances
  In Density Functional Methods: (Part I)}}}\ (\bibinfo  {publisher} {World
  Scientific},\ \bibinfo {year} {1995})\ pp.\ \bibinfo {pages}
  {155--192}\BibitemShut {NoStop}%
\bibitem [{\citenamefont {Albrecht}\ and\ \citenamefont
  {Creighton}(1977)}]{albrecht1977anomalously}%
  \BibitemOpen
  \bibfield  {author} {\bibinfo {author} {\bibfnamefont {M.~G.}\ \bibnamefont
  {Albrecht}}\ and\ \bibinfo {author} {\bibfnamefont {J.~A.}\ \bibnamefont
  {Creighton}},\ }\href@noop {} {\bibfield  {journal} {\bibinfo  {journal} {J.
  Am. Chem. Soc.}\ }\textbf {\bibinfo {volume} {99}},\ \bibinfo {pages} {5215}
  (\bibinfo {year} {1977})}\BibitemShut {NoStop}%
\bibitem [{\citenamefont {Jeanmaire}\ and\ \citenamefont
  {Van~Duyne}(1977)}]{jeanmaire1977surface}%
  \BibitemOpen
  \bibfield  {author} {\bibinfo {author} {\bibfnamefont {D.~L.}\ \bibnamefont
  {Jeanmaire}}\ and\ \bibinfo {author} {\bibfnamefont {R.~P.}\ \bibnamefont
  {Van~Duyne}},\ }\href@noop {} {\bibfield  {journal} {\bibinfo  {journal} {J.
  Electroanal. Chem. Interf. Electrochem.}\ }\textbf {\bibinfo {volume} {84}},\
  \bibinfo {pages} {1} (\bibinfo {year} {1977})}\BibitemShut {NoStop}%
\bibitem [{\citenamefont {Campion}\ and\ \citenamefont
  {Kambhampati}(1998)}]{campion1998surface}%
  \BibitemOpen
  \bibfield  {author} {\bibinfo {author} {\bibfnamefont {A.}~\bibnamefont
  {Campion}}\ and\ \bibinfo {author} {\bibfnamefont {P.}~\bibnamefont
  {Kambhampati}},\ }\href@noop {} {\bibfield  {journal} {\bibinfo  {journal}
  {Chem. Soc. Rev.}\ }\textbf {\bibinfo {volume} {27}},\ \bibinfo {pages} {241}
  (\bibinfo {year} {1998})}\BibitemShut {NoStop}%
\bibitem [{\citenamefont {Talley}\ \emph {et~al.}(2005)\citenamefont {Talley},
  \citenamefont {Jackson}, \citenamefont {Oubre}, \citenamefont {Grady},
  \citenamefont {Hollars}, \citenamefont {Lane}, \citenamefont {Huser},
  \citenamefont {Nordlander},\ and\ \citenamefont {Halas}}]{Talley2005Surface}%
  \BibitemOpen
  \bibfield  {author} {\bibinfo {author} {\bibfnamefont {C.~E.}\ \bibnamefont
  {Talley}}, \bibinfo {author} {\bibfnamefont {J.~B.}\ \bibnamefont {Jackson}},
  \bibinfo {author} {\bibfnamefont {C.}~\bibnamefont {Oubre}}, \bibinfo
  {author} {\bibfnamefont {N.~K.}\ \bibnamefont {Grady}}, \bibinfo {author}
  {\bibfnamefont {C.~W.}\ \bibnamefont {Hollars}}, \bibinfo {author}
  {\bibfnamefont {S.~M.}\ \bibnamefont {Lane}}, \bibinfo {author}
  {\bibfnamefont {T.~R.}\ \bibnamefont {Huser}}, \bibinfo {author}
  {\bibfnamefont {P.~J.}\ \bibnamefont {Nordlander}}, \ and\ \bibinfo {author}
  {\bibfnamefont {N.~J.}\ \bibnamefont {Halas}},\ }\href
  {https://api.semanticscholar.org/CorpusID:34302236} {\bibfield  {journal}
  {\bibinfo  {journal} {Nano Lett.}\ }\textbf {\bibinfo {volume} {5 8}},\
  \bibinfo {pages} {1569} (\bibinfo {year} {2005})}\BibitemShut {NoStop}%
\bibitem [{\citenamefont {Alvarez-Puebla}, \citenamefont {Liz-Marz{\'a}n},\
  and\ \citenamefont {Garc{\'i}a~de Abajo}(2010)}]{alvarez2010light}%
  \BibitemOpen
  \bibfield  {author} {\bibinfo {author} {\bibfnamefont {R.}~\bibnamefont
  {Alvarez-Puebla}}, \bibinfo {author} {\bibfnamefont {L.~M.}\ \bibnamefont
  {Liz-Marz{\'a}n}}, \ and\ \bibinfo {author} {\bibfnamefont {F.~J.}\
  \bibnamefont {Garc{\'i}a~de Abajo}},\ }\href@noop {} {\bibfield  {journal}
  {\bibinfo  {journal} {J. Phys. Chem. Lett.}\ }\textbf {\bibinfo {volume}
  {1}},\ \bibinfo {pages} {2428} (\bibinfo {year} {2010})}\BibitemShut
  {NoStop}%
\bibitem [{\citenamefont {Le~Ru}\ and\ \citenamefont
  {Etchegoin}(2008)}]{le2008principles}%
  \BibitemOpen
  \bibfield  {author} {\bibinfo {author} {\bibfnamefont {E.}~\bibnamefont
  {Le~Ru}}\ and\ \bibinfo {author} {\bibfnamefont {P.}~\bibnamefont
  {Etchegoin}},\ }\href@noop {} {\emph {\bibinfo {title} {Principles of
  Surface-Enhanced Raman Spectroscopy: and related plasmonic effects}}}\
  (\bibinfo  {publisher} {Elsevier},\ \bibinfo {year} {2008})\BibitemShut
  {NoStop}%
\bibitem [{\citenamefont {Sharma}\ \emph {et~al.}(2012)\citenamefont {Sharma},
  \citenamefont {Frontiera}, \citenamefont {Henry}, \citenamefont {Ringe},\
  and\ \citenamefont {{Van Duyne}}}]{SHARMA201216}%
  \BibitemOpen
  \bibfield  {author} {\bibinfo {author} {\bibfnamefont {B.}~\bibnamefont
  {Sharma}}, \bibinfo {author} {\bibfnamefont {R.~R.}\ \bibnamefont
  {Frontiera}}, \bibinfo {author} {\bibfnamefont {A.-I.}\ \bibnamefont
  {Henry}}, \bibinfo {author} {\bibfnamefont {E.}~\bibnamefont {Ringe}}, \ and\
  \bibinfo {author} {\bibfnamefont {R.~P.}\ \bibnamefont {{Van Duyne}}},\
  }\href {\doibase https://doi.org/10.1016/S1369-7021(12)70017-2} {\bibfield
  {journal} {\bibinfo  {journal} {Mater. Today}\ }\textbf {\bibinfo {volume}
  {15}},\ \bibinfo {pages} {16} (\bibinfo {year} {2012})}\BibitemShut {NoStop}%
\bibitem [{\citenamefont {Han}\ \emph {et~al.}(2022)\citenamefont {Han},
  \citenamefont {Rodriguez}, \citenamefont {Haynes}, \citenamefont {Ozaki},\
  and\ \citenamefont {Zhao}}]{han2022surface}%
  \BibitemOpen
  \bibfield  {author} {\bibinfo {author} {\bibfnamefont {X.~X.}\ \bibnamefont
  {Han}}, \bibinfo {author} {\bibfnamefont {R.~S.}\ \bibnamefont {Rodriguez}},
  \bibinfo {author} {\bibfnamefont {C.~L.}\ \bibnamefont {Haynes}}, \bibinfo
  {author} {\bibfnamefont {Y.}~\bibnamefont {Ozaki}}, \ and\ \bibinfo {author}
  {\bibfnamefont {B.}~\bibnamefont {Zhao}},\ }\href@noop {} {\bibfield
  {journal} {\bibinfo  {journal} {Nat. Rev. Methods Primers}\ }\textbf
  {\bibinfo {volume} {1}},\ \bibinfo {pages} {1} (\bibinfo {year}
  {2022})}\BibitemShut {NoStop}%
\bibitem [{\citenamefont {Lai}\ \emph {et~al.}(2018)\citenamefont {Lai},
  \citenamefont {Xu}, \citenamefont {Zhang},\ and\ \citenamefont
  {Wang}}]{lai2018recent}%
  \BibitemOpen
  \bibfield  {author} {\bibinfo {author} {\bibfnamefont {H.}~\bibnamefont
  {Lai}}, \bibinfo {author} {\bibfnamefont {F.}~\bibnamefont {Xu}}, \bibinfo
  {author} {\bibfnamefont {Y.}~\bibnamefont {Zhang}}, \ and\ \bibinfo {author}
  {\bibfnamefont {L.}~\bibnamefont {Wang}},\ }\href@noop {} {\bibfield
  {journal} {\bibinfo  {journal} {J. Mater. Chem. B}\ }\textbf {\bibinfo
  {volume} {6}},\ \bibinfo {pages} {4008} (\bibinfo {year} {2018})}\BibitemShut
  {NoStop}%
\bibitem [{\citenamefont {Zhang}\ \emph {et~al.}(2020)\citenamefont {Zhang},
  \citenamefont {Duan}, \citenamefont {Radjenovic}, \citenamefont {Tian},\ and\
  \citenamefont {Li}}]{zhang2020core}%
  \BibitemOpen
  \bibfield  {author} {\bibinfo {author} {\bibfnamefont {H.}~\bibnamefont
  {Zhang}}, \bibinfo {author} {\bibfnamefont {S.}~\bibnamefont {Duan}},
  \bibinfo {author} {\bibfnamefont {P.~M.}\ \bibnamefont {Radjenovic}},
  \bibinfo {author} {\bibfnamefont {Z.-Q.}\ \bibnamefont {Tian}}, \ and\
  \bibinfo {author} {\bibfnamefont {J.-F.}\ \bibnamefont {Li}},\ }\href
  {\doibase 10.1021/acs.accounts.9b00545} {\bibfield  {journal} {\bibinfo
  {journal} {Acc. Chem. Res.}\ }\textbf {\bibinfo {volume} {53}},\ \bibinfo
  {pages} {729} (\bibinfo {year} {2020})}\BibitemShut {NoStop}%
\bibitem [{\citenamefont {Su}\ \emph {et~al.}(2021)\citenamefont {Su},
  \citenamefont {Feng}, \citenamefont {Wu}, \citenamefont {juan Sun},\ and\
  \citenamefont {Ren}}]{Su2021Recent}%
  \BibitemOpen
  \bibfield  {author} {\bibinfo {author} {\bibfnamefont {H.}~\bibnamefont
  {Su}}, \bibinfo {author} {\bibfnamefont {H.-S.}\ \bibnamefont {Feng}},
  \bibinfo {author} {\bibfnamefont {X.}~\bibnamefont {Wu}}, \bibinfo {author}
  {\bibfnamefont {J.}~\bibnamefont {juan Sun}}, \ and\ \bibinfo {author}
  {\bibfnamefont {B.}~\bibnamefont {Ren}},\ }\href
  {https://api.semanticscholar.org/CorpusID:237401335} {\bibfield  {journal}
  {\bibinfo  {journal} {Nanoscale}\ }\textbf {\bibinfo {volume} {13 33}},\
  \bibinfo {pages} {13962} (\bibinfo {year} {2021})}\BibitemShut {NoStop}%
\bibitem [{\citenamefont {Zhao}\ \emph {et~al.}(2020)\citenamefont {Zhao},
  \citenamefont {Campbell}, \citenamefont {Wallace}, \citenamefont {Claing},
  \citenamefont {Bazuin},\ and\ \citenamefont {Masson}}]{zhao2020branched}%
  \BibitemOpen
  \bibfield  {author} {\bibinfo {author} {\bibfnamefont {X.}~\bibnamefont
  {Zhao}}, \bibinfo {author} {\bibfnamefont {S.}~\bibnamefont {Campbell}},
  \bibinfo {author} {\bibfnamefont {G.~Q.}\ \bibnamefont {Wallace}}, \bibinfo
  {author} {\bibfnamefont {A.}~\bibnamefont {Claing}}, \bibinfo {author}
  {\bibfnamefont {C.~G.}\ \bibnamefont {Bazuin}}, \ and\ \bibinfo {author}
  {\bibfnamefont {J.-F.}\ \bibnamefont {Masson}},\ }\href {\doibase
  10.1021/acssensors.0c00784} {\bibfield  {journal} {\bibinfo  {journal} {ACS
  Sens.}\ }\textbf {\bibinfo {volume} {5}},\ \bibinfo {pages} {2155} (\bibinfo
  {year} {2020})}\BibitemShut {NoStop}%
\bibitem [{\citenamefont {Fang}\ \emph {et~al.}(2023)\citenamefont {Fang},
  \citenamefont {Pan}, \citenamefont {Liu}, \citenamefont {Jiang},
  \citenamefont {Ye}, \citenamefont {Ji}, \citenamefont {Wang},\ and\
  \citenamefont {Xia}}]{Fang2023Surface}%
  \BibitemOpen
  \bibfield  {author} {\bibinfo {author} {\bibfnamefont {L.}~\bibnamefont
  {Fang}}, \bibinfo {author} {\bibfnamefont {X.-T.}\ \bibnamefont {Pan}},
  \bibinfo {author} {\bibfnamefont {K.}~\bibnamefont {Liu}}, \bibinfo {author}
  {\bibfnamefont {D.}~\bibnamefont {Jiang}}, \bibinfo {author} {\bibfnamefont
  {D.}~\bibnamefont {Ye}}, \bibinfo {author} {\bibfnamefont {L.-N.}\
  \bibnamefont {Ji}}, \bibinfo {author} {\bibfnamefont {K.}~\bibnamefont
  {Wang}}, \ and\ \bibinfo {author} {\bibfnamefont {X.}~\bibnamefont {Xia}},\
  }\href {https://api.semanticscholar.org/CorpusID:258214910} {\bibfield
  {journal} {\bibinfo  {journal} {ACS Appl. Mater. Interfaces}\ } (\bibinfo
  {year} {2023})}\BibitemShut {NoStop}%
\bibitem [{\citenamefont {Troncoso-Afonso}\ \emph {et~al.}(2024)\citenamefont
  {Troncoso-Afonso}, \citenamefont {Vinnacombe-Willson}, \citenamefont
  {Garc{\'i}a‐Astrain},\ and\ \citenamefont
  {Liz‐Marz{\'a}n}}]{Troncoso2024SERS}%
  \BibitemOpen
  \bibfield  {author} {\bibinfo {author} {\bibfnamefont {L.}~\bibnamefont
  {Troncoso-Afonso}}, \bibinfo {author} {\bibfnamefont {G.~A.}\ \bibnamefont
  {Vinnacombe-Willson}}, \bibinfo {author} {\bibfnamefont {C.}~\bibnamefont
  {Garc{\'i}a‐Astrain}}, \ and\ \bibinfo {author} {\bibfnamefont {L.~M.}\
  \bibnamefont {Liz‐Marz{\'a}n}},\ }\href
  {https://api.semanticscholar.org/CorpusID:269087223} {\bibfield  {journal}
  {\bibinfo  {journal} {Chem. Soc. Rev.}\ } (\bibinfo {year}
  {2024})}\BibitemShut {NoStop}%
\bibitem [{\citenamefont {Perumal}\ \emph {et~al.}(2021)\citenamefont
  {Perumal}, \citenamefont {Wang}, \citenamefont {Attia}, \citenamefont
  {Dinish},\ and\ \citenamefont {Olivo}}]{Perumal2021Towards}%
  \BibitemOpen
  \bibfield  {author} {\bibinfo {author} {\bibfnamefont {J.}~\bibnamefont
  {Perumal}}, \bibinfo {author} {\bibfnamefont {Y.}~\bibnamefont {Wang}},
  \bibinfo {author} {\bibfnamefont {A.~B.~E.}\ \bibnamefont {Attia}}, \bibinfo
  {author} {\bibfnamefont {U.~S.}\ \bibnamefont {Dinish}}, \ and\ \bibinfo
  {author} {\bibfnamefont {M.}~\bibnamefont {Olivo}},\ }\href
  {https://api.semanticscholar.org/CorpusID:230784506} {\bibfield  {journal}
  {\bibinfo  {journal} {Nanoscale}\ } (\bibinfo {year} {2021})}\BibitemShut
  {NoStop}%
\bibitem [{\citenamefont {Taheri-Ledari}\ \emph {et~al.}(2023)\citenamefont
  {Taheri-Ledari}, \citenamefont {Ganjali}, \citenamefont {Zarei-Shokat},
  \citenamefont {Dinmohammadi}, \citenamefont {Asl}, \citenamefont {Emami},
  \citenamefont {Mojtabapour}, \citenamefont {Rashvandi}, \citenamefont
  {Kashtiaray}, \citenamefont {Jalali},\ and\ \citenamefont
  {Maleki}}]{Taheri2023Plasmonic}%
  \BibitemOpen
  \bibfield  {author} {\bibinfo {author} {\bibfnamefont {R.}~\bibnamefont
  {Taheri-Ledari}}, \bibinfo {author} {\bibfnamefont {F.}~\bibnamefont
  {Ganjali}}, \bibinfo {author} {\bibfnamefont {S.}~\bibnamefont
  {Zarei-Shokat}}, \bibinfo {author} {\bibfnamefont {R.}~\bibnamefont
  {Dinmohammadi}}, \bibinfo {author} {\bibfnamefont {F.~R.}\ \bibnamefont
  {Asl}}, \bibinfo {author} {\bibfnamefont {A.}~\bibnamefont {Emami}}, \bibinfo
  {author} {\bibfnamefont {Z.}~\bibnamefont {Mojtabapour}}, \bibinfo {author}
  {\bibfnamefont {Z.}~\bibnamefont {Rashvandi}}, \bibinfo {author}
  {\bibfnamefont {A.}~\bibnamefont {Kashtiaray}}, \bibinfo {author}
  {\bibfnamefont {F.}~\bibnamefont {Jalali}}, \ and\ \bibinfo {author}
  {\bibfnamefont {A.}~\bibnamefont {Maleki}},\ }\href
  {https://api.semanticscholar.org/CorpusID:265487357} {\bibfield  {journal}
  {\bibinfo  {journal} {Nanoscale Adv.}\ }\textbf {\bibinfo {volume} {5}},\
  \bibinfo {pages} {6768 } (\bibinfo {year} {2023})}\BibitemShut {NoStop}%
\bibitem [{\citenamefont {Jackson}(1999)}]{jackson1999classical}%
  \BibitemOpen
  \bibfield  {author} {\bibinfo {author} {\bibfnamefont {J.~D.}\ \bibnamefont
  {Jackson}},\ }\enquote {\bibinfo {title} {Classical electrodynamics},}\ \
  (\bibinfo  {publisher} {John Wiley \& Sons, Ltd},\ \bibinfo {year}
  {1999})\BibitemShut {NoStop}%
\bibitem [{\citenamefont {Pinchuk}, \citenamefont {Von~Plessen},\ and\
  \citenamefont {Kreibig}(2004)}]{pinchuk2004influence}%
  \BibitemOpen
  \bibfield  {author} {\bibinfo {author} {\bibfnamefont {A.}~\bibnamefont
  {Pinchuk}}, \bibinfo {author} {\bibfnamefont {G.}~\bibnamefont
  {Von~Plessen}}, \ and\ \bibinfo {author} {\bibfnamefont {U.}~\bibnamefont
  {Kreibig}},\ }\href@noop {} {\bibfield  {journal} {\bibinfo  {journal} {J.
  Phys. D: Appl. Phys.}\ }\textbf {\bibinfo {volume} {37}},\ \bibinfo {pages}
  {3133} (\bibinfo {year} {2004})}\BibitemShut {NoStop}%
\bibitem [{\citenamefont {Pinchuk}, \citenamefont {Kreibig},\ and\
  \citenamefont {Hilger}(2004)}]{pinchuk2004optical}%
  \BibitemOpen
  \bibfield  {author} {\bibinfo {author} {\bibfnamefont {A.}~\bibnamefont
  {Pinchuk}}, \bibinfo {author} {\bibfnamefont {U.}~\bibnamefont {Kreibig}}, \
  and\ \bibinfo {author} {\bibfnamefont {A.}~\bibnamefont {Hilger}},\
  }\href@noop {} {\bibfield  {journal} {\bibinfo  {journal} {Surf. Sci.}\
  }\textbf {\bibinfo {volume} {557}},\ \bibinfo {pages} {269} (\bibinfo {year}
  {2004})}\BibitemShut {NoStop}%
\bibitem [{\citenamefont {Balamurugan}\ and\ \citenamefont
  {Maruyama}(2005)}]{balamurugan2005evidence}%
  \BibitemOpen
  \bibfield  {author} {\bibinfo {author} {\bibfnamefont {B.}~\bibnamefont
  {Balamurugan}}\ and\ \bibinfo {author} {\bibfnamefont {T.}~\bibnamefont
  {Maruyama}},\ }\href@noop {} {\bibfield  {journal} {\bibinfo  {journal}
  {Appl. Phys. Lett.}\ }\textbf {\bibinfo {volume} {87}},\ \bibinfo {pages}
  {143105} (\bibinfo {year} {2005})}\BibitemShut {NoStop}%
\bibitem [{\citenamefont {Giovannini}\ \emph
  {et~al.}(2019{\natexlab{b}})\citenamefont {Giovannini}, \citenamefont
  {Puglisi}, \citenamefont {Ambrosetti},\ and\ \citenamefont
  {Cappelli}}]{giovannini2019polarizable}%
  \BibitemOpen
  \bibfield  {author} {\bibinfo {author} {\bibfnamefont {T.}~\bibnamefont
  {Giovannini}}, \bibinfo {author} {\bibfnamefont {A.}~\bibnamefont {Puglisi}},
  \bibinfo {author} {\bibfnamefont {M.}~\bibnamefont {Ambrosetti}}, \ and\
  \bibinfo {author} {\bibfnamefont {C.}~\bibnamefont {Cappelli}},\ }\href@noop
  {} {\bibfield  {journal} {\bibinfo  {journal} {J. Chem. Theory Comput.}\
  }\textbf {\bibinfo {volume} {15}},\ \bibinfo {pages} {2233} (\bibinfo {year}
  {2019}{\natexlab{b}})}\BibitemShut {NoStop}%
\bibitem [{\citenamefont {Mayer}(2007)}]{mayer2007formulation}%
  \BibitemOpen
  \bibfield  {author} {\bibinfo {author} {\bibfnamefont {A.}~\bibnamefont
  {Mayer}},\ }\href@noop {} {\bibfield  {journal} {\bibinfo  {journal} {Phys.
  Rev. B}\ }\textbf {\bibinfo {volume} {75}},\ \bibinfo {pages} {045407}
  (\bibinfo {year} {2007})}\BibitemShut {NoStop}%
\bibitem [{\citenamefont {Fuchs}(1975)}]{fuchs1975}%
  \BibitemOpen
  \bibfield  {author} {\bibinfo {author} {\bibfnamefont {R.}~\bibnamefont
  {Fuchs}},\ }\href {\doibase 10.1103/PhysRevB.11.1732} {\bibfield  {journal}
  {\bibinfo  {journal} {Phys. Rev. B}\ }\textbf {\bibinfo {volume} {11}},\
  \bibinfo {pages} {1732} (\bibinfo {year} {1975})}\BibitemShut {NoStop}%
\bibitem [{\citenamefont {Tomasi}, \citenamefont {Mennucci},\ and\
  \citenamefont {Cammi}(2005)}]{tomasi2005quantum}%
  \BibitemOpen
  \bibfield  {author} {\bibinfo {author} {\bibfnamefont {J.}~\bibnamefont
  {Tomasi}}, \bibinfo {author} {\bibfnamefont {B.}~\bibnamefont {Mennucci}}, \
  and\ \bibinfo {author} {\bibfnamefont {R.}~\bibnamefont {Cammi}},\ }\href
  {\doibase 10.1021/cr9904009} {\bibfield  {journal} {\bibinfo  {journal}
  {Chem. Rev.}\ }\textbf {\bibinfo {volume} {105}},\ \bibinfo {pages} {2999}
  (\bibinfo {year} {2005})}\BibitemShut {NoStop}%
\bibitem [{\citenamefont {Vukovic}, \citenamefont {Corni},\ and\ \citenamefont
  {Mennucci}(2009)}]{vukovic2009fluorescence}%
  \BibitemOpen
  \bibfield  {author} {\bibinfo {author} {\bibfnamefont {S.}~\bibnamefont
  {Vukovic}}, \bibinfo {author} {\bibfnamefont {S.}~\bibnamefont {Corni}}, \
  and\ \bibinfo {author} {\bibfnamefont {B.}~\bibnamefont {Mennucci}},\
  }\href@noop {} {\bibfield  {journal} {\bibinfo  {journal} {J. Phys. Chem. C}\
  }\textbf {\bibinfo {volume} {113}},\ \bibinfo {pages} {121} (\bibinfo {year}
  {2009})}\BibitemShut {NoStop}%
\bibitem [{\citenamefont {Grobas~Illobre}\ \emph {et~al.}(2024)\citenamefont
  {Grobas~Illobre}, \citenamefont {Lafiosca}, \citenamefont {Guidone},
  \citenamefont {Mazza}, \citenamefont {Giovannini},\ and\ \citenamefont
  {Cappelli}}]{grobas2024multiscale}%
  \BibitemOpen
  \bibfield  {author} {\bibinfo {author} {\bibfnamefont {P.}~\bibnamefont
  {Grobas~Illobre}}, \bibinfo {author} {\bibfnamefont {P.}~\bibnamefont
  {Lafiosca}}, \bibinfo {author} {\bibfnamefont {T.}~\bibnamefont {Guidone}},
  \bibinfo {author} {\bibfnamefont {F.}~\bibnamefont {Mazza}}, \bibinfo
  {author} {\bibfnamefont {T.}~\bibnamefont {Giovannini}}, \ and\ \bibinfo
  {author} {\bibfnamefont {C.}~\bibnamefont {Cappelli}},\ }\href {\doibase
  10.1039/D4NA00080C} {\bibfield  {journal} {\bibinfo  {journal} {Nanoscale
  Adv.}\ } (\bibinfo {year} {2024}),\ 10.1039/D4NA00080C}\BibitemShut {NoStop}%
\bibitem [{\citenamefont {Helgaker}, \citenamefont {Jørgensen},\ and\
  \citenamefont {Olsen}(2000)}]{molecularelectronicstructuretheory-ch9}%
  \BibitemOpen
  \bibfield  {author} {\bibinfo {author} {\bibfnamefont {T.}~\bibnamefont
  {Helgaker}}, \bibinfo {author} {\bibfnamefont {P.}~\bibnamefont
  {Jørgensen}}, \ and\ \bibinfo {author} {\bibfnamefont {J.}~\bibnamefont
  {Olsen}},\ }\enquote {\bibinfo {title} {Molecular integral evaluation},}\ in\
  \href {\doibase https://doi.org/10.1002/9781119019572.ch9} {\emph {\bibinfo
  {booktitle} {Molecular Electronic‐Structure Theory}}}\ (\bibinfo
  {publisher} {John Wiley \& Sons, Ltd},\ \bibinfo {year} {2000})\
  Chap.~\bibinfo {chapter} {9}, pp.\ \bibinfo {pages} {336--432}\BibitemShut
  {NoStop}%
\bibitem [{\citenamefont {Warshel}\ and\ \citenamefont
  {Levitt}(1976)}]{warshel1976theoretical}%
  \BibitemOpen
  \bibfield  {author} {\bibinfo {author} {\bibfnamefont {A.}~\bibnamefont
  {Warshel}}\ and\ \bibinfo {author} {\bibfnamefont {M.}~\bibnamefont
  {Levitt}},\ }\href@noop {} {\bibfield  {journal} {\bibinfo  {journal} {J.
  Mol. Biol.}\ }\textbf {\bibinfo {volume} {103}},\ \bibinfo {pages} {227}
  (\bibinfo {year} {1976})}\BibitemShut {NoStop}%
\bibitem [{\citenamefont {Lin}\ and\ \citenamefont
  {Truhlar}(2007)}]{lin2007qm}%
  \BibitemOpen
  \bibfield  {author} {\bibinfo {author} {\bibfnamefont {H.}~\bibnamefont
  {Lin}}\ and\ \bibinfo {author} {\bibfnamefont {D.~G.}\ \bibnamefont
  {Truhlar}},\ }\href@noop {} {\bibfield  {journal} {\bibinfo  {journal}
  {Theor. Chem. Acc.}\ }\textbf {\bibinfo {volume} {117}},\ \bibinfo {pages}
  {185} (\bibinfo {year} {2007})}\BibitemShut {NoStop}%
\bibitem [{\citenamefont {Senn}\ and\ \citenamefont
  {Thiel}(2009)}]{senn2009qm}%
  \BibitemOpen
  \bibfield  {author} {\bibinfo {author} {\bibfnamefont {H.~M.}\ \bibnamefont
  {Senn}}\ and\ \bibinfo {author} {\bibfnamefont {W.}~\bibnamefont {Thiel}},\
  }\href@noop {} {\bibfield  {journal} {\bibinfo  {journal} {Angew. Chem. Int.
  Ed.}\ }\textbf {\bibinfo {volume} {48}},\ \bibinfo {pages} {1198} (\bibinfo
  {year} {2009})}\BibitemShut {NoStop}%
\bibitem [{\citenamefont {Guido}\ \emph {et~al.}(2020)\citenamefont {Guido},
  \citenamefont {Rosa}, \citenamefont {Cammi},\ and\ \citenamefont
  {Corni}}]{guido2020open}%
  \BibitemOpen
  \bibfield  {author} {\bibinfo {author} {\bibfnamefont {C.~A.}\ \bibnamefont
  {Guido}}, \bibinfo {author} {\bibfnamefont {M.}~\bibnamefont {Rosa}},
  \bibinfo {author} {\bibfnamefont {R.}~\bibnamefont {Cammi}}, \ and\ \bibinfo
  {author} {\bibfnamefont {S.}~\bibnamefont {Corni}},\ }\href@noop {}
  {\bibfield  {journal} {\bibinfo  {journal} {J. Chem. Phys.}\ }\textbf
  {\bibinfo {volume} {152}},\ \bibinfo {pages} {174114} (\bibinfo {year}
  {2020})}\BibitemShut {NoStop}%
\bibitem [{\citenamefont {Corni}, \citenamefont {Pipolo},\ and\ \citenamefont
  {Cammi}(2015)}]{corni2015equation}%
  \BibitemOpen
  \bibfield  {author} {\bibinfo {author} {\bibfnamefont {S.}~\bibnamefont
  {Corni}}, \bibinfo {author} {\bibfnamefont {S.}~\bibnamefont {Pipolo}}, \
  and\ \bibinfo {author} {\bibfnamefont {R.}~\bibnamefont {Cammi}},\
  }\href@noop {} {\bibfield  {journal} {\bibinfo  {journal} {J. Phys. Chem. A}\
  }\textbf {\bibinfo {volume} {119}},\ \bibinfo {pages} {5405} (\bibinfo {year}
  {2015})}\BibitemShut {NoStop}%
\bibitem [{\citenamefont {Norman}, \citenamefont {Ruud},\ and\ \citenamefont
  {Saue}(2018)}]{norman2018principles}%
  \BibitemOpen
  \bibfield  {author} {\bibinfo {author} {\bibfnamefont {P.}~\bibnamefont
  {Norman}}, \bibinfo {author} {\bibfnamefont {K.}~\bibnamefont {Ruud}}, \ and\
  \bibinfo {author} {\bibfnamefont {T.}~\bibnamefont {Saue}},\ }\href@noop {}
  {\emph {\bibinfo {title} {Principles and practices of molecular properties:
  Theory, modeling, and simulations}}}\ (\bibinfo  {publisher} {John Wiley \&
  Sons},\ \bibinfo {year} {2018})\BibitemShut {NoStop}%
\bibitem [{\citenamefont {Baerends}\ and\ \citenamefont {et.
  al}(2020)}]{ams2020}%
  \BibitemOpen
  \bibfield  {author} {\bibinfo {author} {\bibfnamefont {E.}~\bibnamefont
  {Baerends}}\ and\ \bibinfo {author} {\bibnamefont {et. al}},\ }\href@noop {}
  {\enquote {\bibinfo {title} {Adf {\em(version 2020.x)}},}\ } (\bibinfo {year}
  {2020}),\ \bibinfo {note} {theoretical Chemistry, Vrije Universiteit,
  Amsterdam, The Netherlands, http://www.scm.com}\BibitemShut {NoStop}%
\bibitem [{\citenamefont {Nicoli}, \citenamefont {Giovannini},\ and\
  \citenamefont {Cappelli}(2022)}]{nicoli2022assessing}%
  \BibitemOpen
  \bibfield  {author} {\bibinfo {author} {\bibfnamefont {L.}~\bibnamefont
  {Nicoli}}, \bibinfo {author} {\bibfnamefont {T.}~\bibnamefont {Giovannini}},
  \ and\ \bibinfo {author} {\bibfnamefont {C.}~\bibnamefont {Cappelli}},\
  }\href@noop {} {\bibfield  {journal} {\bibinfo  {journal} {J. Chem. Phys.}\ }
  (\bibinfo {year} {2022})}\BibitemShut {NoStop}%
\bibitem [{\citenamefont {Jensen}, \citenamefont {Autschbach},\ and\
  \citenamefont {Schatz}(2005)}]{jensen2005finite}%
  \BibitemOpen
  \bibfield  {author} {\bibinfo {author} {\bibfnamefont {L.}~\bibnamefont
  {Jensen}}, \bibinfo {author} {\bibfnamefont {J.}~\bibnamefont {Autschbach}},
  \ and\ \bibinfo {author} {\bibfnamefont {G.~C.}\ \bibnamefont {Schatz}},\
  }\href@noop {} {\bibfield  {journal} {\bibinfo  {journal} {J. Chem. Phys.}\
  }\textbf {\bibinfo {volume} {122}},\ \bibinfo {pages} {224115} (\bibinfo
  {year} {2005})}\BibitemShut {NoStop}%
\bibitem [{\citenamefont {Payton}\ \emph {et~al.}(2014)\citenamefont {Payton},
  \citenamefont {Morton}, \citenamefont {Moore},\ and\ \citenamefont
  {Jensen}}]{payton2014hybrid}%
  \BibitemOpen
  \bibfield  {author} {\bibinfo {author} {\bibfnamefont {J.~L.}\ \bibnamefont
  {Payton}}, \bibinfo {author} {\bibfnamefont {S.~M.}\ \bibnamefont {Morton}},
  \bibinfo {author} {\bibfnamefont {J.~E.}\ \bibnamefont {Moore}}, \ and\
  \bibinfo {author} {\bibfnamefont {L.}~\bibnamefont {Jensen}},\ }\href@noop {}
  {\bibfield  {journal} {\bibinfo  {journal} {Acc. Chem. Res.}\ }\textbf
  {\bibinfo {volume} {47}},\ \bibinfo {pages} {88} (\bibinfo {year}
  {2014})}\BibitemShut {NoStop}%
\bibitem [{\citenamefont {Placzek}(1934)}]{placzek1934handbuch}%
  \BibitemOpen
  \bibfield  {author} {\bibinfo {author} {\bibfnamefont {G.}~\bibnamefont
  {Placzek}},\ }\href@noop {} {\bibfield  {journal} {\bibinfo  {journal} {Ed.
  G. Marx, Akademische Verlagsgesellschaft, Leipzig}\ } (\bibinfo {year}
  {1934})}\BibitemShut {NoStop}%
\bibitem [{\citenamefont {Placzek}\ and\ \citenamefont
  {Teller}(1933)}]{placzek1933rotationsstruktur}%
  \BibitemOpen
  \bibfield  {author} {\bibinfo {author} {\bibfnamefont {G.}~\bibnamefont
  {Placzek}}\ and\ \bibinfo {author} {\bibfnamefont {E.}~\bibnamefont
  {Teller}},\ }\href@noop {} {\bibfield  {journal} {\bibinfo  {journal} {Z.
  Phys.}\ }\textbf {\bibinfo {volume} {81}},\ \bibinfo {pages} {209} (\bibinfo
  {year} {1933})}\BibitemShut {NoStop}%
\bibitem [{\citenamefont {Jensen}\ \emph {et~al.}(2005)\citenamefont {Jensen},
  \citenamefont {Zhao}, \citenamefont {Autschbach},\ and\ \citenamefont
  {Schatz}}]{jensen2005theory}%
  \BibitemOpen
  \bibfield  {author} {\bibinfo {author} {\bibfnamefont {L.}~\bibnamefont
  {Jensen}}, \bibinfo {author} {\bibfnamefont {L.}~\bibnamefont {Zhao}},
  \bibinfo {author} {\bibfnamefont {J.}~\bibnamefont {Autschbach}}, \ and\
  \bibinfo {author} {\bibfnamefont {G.}~\bibnamefont {Schatz}},\ }\href@noop {}
  {\bibfield  {journal} {\bibinfo  {journal} {J. Chem. Phys.}\ }\textbf
  {\bibinfo {volume} {123}},\ \bibinfo {pages} {174110} (\bibinfo {year}
  {2005})}\BibitemShut {NoStop}%
\bibitem [{\citenamefont {Larsen}\ \emph {et~al.}(2017)\citenamefont {Larsen},
  \citenamefont {Mortensen}, \citenamefont {Blomqvist}, \citenamefont
  {Castelli}, \citenamefont {Christensen}, \citenamefont {Dułak},
  \citenamefont {Friis}, \citenamefont {Groves}, \citenamefont {Hammer},
  \citenamefont {Hargus}, \citenamefont {Hermes}, \citenamefont {Jennings},
  \citenamefont {Jensen}, \citenamefont {Kermode}, \citenamefont {Kitchin},
  \citenamefont {Kolsbjerg}, \citenamefont {Kubal}, \citenamefont {Kaasbjerg},
  \citenamefont {Lysgaard}, \citenamefont {Maronsson}, \citenamefont {Maxson},
  \citenamefont {Olsen}, \citenamefont {Pastewka}, \citenamefont {Peterson},
  \citenamefont {Rostgaard}, \citenamefont {Schiøtz}, \citenamefont {Schütt},
  \citenamefont {Strange}, \citenamefont {Thygesen}, \citenamefont {Vegge},
  \citenamefont {Vilhelmsen}, \citenamefont {Walter}, \citenamefont {Zeng},\
  and\ \citenamefont {Jacobsen}}]{larsen2017atomic}%
  \BibitemOpen
  \bibfield  {author} {\bibinfo {author} {\bibfnamefont {A.~H.}\ \bibnamefont
  {Larsen}}, \bibinfo {author} {\bibfnamefont {J.~J.}\ \bibnamefont
  {Mortensen}}, \bibinfo {author} {\bibfnamefont {J.}~\bibnamefont
  {Blomqvist}}, \bibinfo {author} {\bibfnamefont {I.~E.}\ \bibnamefont
  {Castelli}}, \bibinfo {author} {\bibfnamefont {R.}~\bibnamefont
  {Christensen}}, \bibinfo {author} {\bibfnamefont {M.}~\bibnamefont {Dułak}},
  \bibinfo {author} {\bibfnamefont {J.}~\bibnamefont {Friis}}, \bibinfo
  {author} {\bibfnamefont {M.~N.}\ \bibnamefont {Groves}}, \bibinfo {author}
  {\bibfnamefont {B.}~\bibnamefont {Hammer}}, \bibinfo {author} {\bibfnamefont
  {C.}~\bibnamefont {Hargus}}, \bibinfo {author} {\bibfnamefont {E.~D.}\
  \bibnamefont {Hermes}}, \bibinfo {author} {\bibfnamefont {P.~C.}\
  \bibnamefont {Jennings}}, \bibinfo {author} {\bibfnamefont {P.~B.}\
  \bibnamefont {Jensen}}, \bibinfo {author} {\bibfnamefont {J.}~\bibnamefont
  {Kermode}}, \bibinfo {author} {\bibfnamefont {J.~R.}\ \bibnamefont
  {Kitchin}}, \bibinfo {author} {\bibfnamefont {E.~L.}\ \bibnamefont
  {Kolsbjerg}}, \bibinfo {author} {\bibfnamefont {J.}~\bibnamefont {Kubal}},
  \bibinfo {author} {\bibfnamefont {K.}~\bibnamefont {Kaasbjerg}}, \bibinfo
  {author} {\bibfnamefont {S.}~\bibnamefont {Lysgaard}}, \bibinfo {author}
  {\bibfnamefont {J.~B.}\ \bibnamefont {Maronsson}}, \bibinfo {author}
  {\bibfnamefont {T.}~\bibnamefont {Maxson}}, \bibinfo {author} {\bibfnamefont
  {T.}~\bibnamefont {Olsen}}, \bibinfo {author} {\bibfnamefont
  {L.}~\bibnamefont {Pastewka}}, \bibinfo {author} {\bibfnamefont
  {A.}~\bibnamefont {Peterson}}, \bibinfo {author} {\bibfnamefont
  {C.}~\bibnamefont {Rostgaard}}, \bibinfo {author} {\bibfnamefont
  {J.}~\bibnamefont {Schiøtz}}, \bibinfo {author} {\bibfnamefont
  {O.}~\bibnamefont {Schütt}}, \bibinfo {author} {\bibfnamefont
  {M.}~\bibnamefont {Strange}}, \bibinfo {author} {\bibfnamefont {K.~S.}\
  \bibnamefont {Thygesen}}, \bibinfo {author} {\bibfnamefont {T.}~\bibnamefont
  {Vegge}}, \bibinfo {author} {\bibfnamefont {L.}~\bibnamefont {Vilhelmsen}},
  \bibinfo {author} {\bibfnamefont {M.}~\bibnamefont {Walter}}, \bibinfo
  {author} {\bibfnamefont {Z.}~\bibnamefont {Zeng}}, \ and\ \bibinfo {author}
  {\bibfnamefont {K.~W.}\ \bibnamefont {Jacobsen}},\ }\href@noop {} {\bibfield
  {journal} {\bibinfo  {journal} {J. Phys. Condens. Matter}\ }\textbf {\bibinfo
  {volume} {29}},\ \bibinfo {pages} {273002} (\bibinfo {year}
  {2017})}\BibitemShut {NoStop}%
\bibitem [{\citenamefont {Geuzaine}\ and\ \citenamefont
  {Remacle}(2009)}]{gmsh}%
  \BibitemOpen
  \bibfield  {author} {\bibinfo {author} {\bibfnamefont {C.}~\bibnamefont
  {Geuzaine}}\ and\ \bibinfo {author} {\bibfnamefont {J.-F.}\ \bibnamefont
  {Remacle}},\ }\href {\doibase https://doi.org/10.1002/nme.2579} {\bibfield
  {journal} {\bibinfo  {journal} {Int. J. Numer. Methods Eng.}\ }\textbf
  {\bibinfo {volume} {79}},\ \bibinfo {pages} {1309} (\bibinfo {year}
  {2009})}\BibitemShut {NoStop}%
\bibitem [{\citenamefont {Palik}(1997)}]{palik1997handbook}%
  \BibitemOpen
  \bibfield  {author} {\bibinfo {author} {\bibfnamefont {E.~D.}\ \bibnamefont
  {Palik}},\ }\href@noop {} {\emph {\bibinfo {title} {Handbook of optical
  constants of solids}}}\ (\bibinfo  {publisher} {Elsevier},\ \bibinfo {year}
  {1997})\BibitemShut {NoStop}%
\bibitem [{\citenamefont {Raki\'{c}}\ \emph {et~al.}(1998)\citenamefont
  {Raki\'{c}}, \citenamefont {Djuri\v{s}i\'{c}}, \citenamefont {Elazar},\ and\
  \citenamefont {Majewski}}]{aleksandar1998laser}%
  \BibitemOpen
  \bibfield  {author} {\bibinfo {author} {\bibfnamefont {A.~D.}\ \bibnamefont
  {Raki\'{c}}}, \bibinfo {author} {\bibfnamefont {A.~B.}\ \bibnamefont
  {Djuri\v{s}i\'{c}}}, \bibinfo {author} {\bibfnamefont {J.~M.}\ \bibnamefont
  {Elazar}}, \ and\ \bibinfo {author} {\bibfnamefont {M.~L.}\ \bibnamefont
  {Majewski}},\ }\href {\doibase 10.1364/AO.37.005271} {\bibfield  {journal}
  {\bibinfo  {journal} {Appl. Opt.}\ }\textbf {\bibinfo {volume} {37}},\
  \bibinfo {pages} {5271} (\bibinfo {year} {1998})}\BibitemShut {NoStop}%
\bibitem [{\citenamefont {Johnson}\ and\ \citenamefont
  {Christy}(1972)}]{johnson1972optical}%
  \BibitemOpen
  \bibfield  {author} {\bibinfo {author} {\bibfnamefont {P.~B.}\ \bibnamefont
  {Johnson}}\ and\ \bibinfo {author} {\bibfnamefont {R.-W.}\ \bibnamefont
  {Christy}},\ }\href@noop {} {\bibfield  {journal} {\bibinfo  {journal} {Phys.
  Rev. B}\ }\textbf {\bibinfo {volume} {6}},\ \bibinfo {pages} {4370} (\bibinfo
  {year} {1972})}\BibitemShut {NoStop}%
\bibitem [{\citenamefont {Van~Gisbergen}, \citenamefont {Snijders},\ and\
  \citenamefont {Baerends}(1995)}]{van1995density}%
  \BibitemOpen
  \bibfield  {author} {\bibinfo {author} {\bibfnamefont {S.}~\bibnamefont
  {Van~Gisbergen}}, \bibinfo {author} {\bibfnamefont {J.}~\bibnamefont
  {Snijders}}, \ and\ \bibinfo {author} {\bibfnamefont {E.}~\bibnamefont
  {Baerends}},\ }\href@noop {} {\bibfield  {journal} {\bibinfo  {journal} {J.
  Chem. Phys.}\ }\textbf {\bibinfo {volume} {103}},\ \bibinfo {pages} {9347}
  (\bibinfo {year} {1995})}\BibitemShut {NoStop}%
\bibitem [{\citenamefont {Van~Gisbergen}, \citenamefont {Snijders},\ and\
  \citenamefont {Baerends}(1999)}]{van1999implementation}%
  \BibitemOpen
  \bibfield  {author} {\bibinfo {author} {\bibfnamefont {S.}~\bibnamefont
  {Van~Gisbergen}}, \bibinfo {author} {\bibfnamefont {J.}~\bibnamefont
  {Snijders}}, \ and\ \bibinfo {author} {\bibfnamefont {E.}~\bibnamefont
  {Baerends}},\ }\href@noop {} {\bibfield  {journal} {\bibinfo  {journal}
  {Comput. Phys. Commun.}\ }\textbf {\bibinfo {volume} {118}},\ \bibinfo
  {pages} {119} (\bibinfo {year} {1999})}\BibitemShut {NoStop}%
\bibitem [{\citenamefont {Fan}\ and\ \citenamefont
  {Ziegler}(1992{\natexlab{a}})}]{fan1992application}%
  \BibitemOpen
  \bibfield  {author} {\bibinfo {author} {\bibfnamefont {L.}~\bibnamefont
  {Fan}}\ and\ \bibinfo {author} {\bibfnamefont {T.}~\bibnamefont {Ziegler}},\
  }\href@noop {} {\bibfield  {journal} {\bibinfo  {journal} {J. Chem. Phys.}\
  }\textbf {\bibinfo {volume} {96}},\ \bibinfo {pages} {9005} (\bibinfo {year}
  {1992}{\natexlab{a}})}\BibitemShut {NoStop}%
\bibitem [{\citenamefont {Fan}\ and\ \citenamefont
  {Ziegler}(1992{\natexlab{b}})}]{fan1992application2}%
  \BibitemOpen
  \bibfield  {author} {\bibinfo {author} {\bibfnamefont {L.}~\bibnamefont
  {Fan}}\ and\ \bibinfo {author} {\bibfnamefont {T.}~\bibnamefont {Ziegler}},\
  }\href@noop {} {\bibfield  {journal} {\bibinfo  {journal} {J. Phys. Chem.}\
  }\textbf {\bibinfo {volume} {96}},\ \bibinfo {pages} {6937} (\bibinfo {year}
  {1992}{\natexlab{b}})}\BibitemShut {NoStop}%
\bibitem [{\citenamefont {Van~Gisbergen}, \citenamefont {Snijders},\ and\
  \citenamefont {Baerends}(1996)}]{van1996application}%
  \BibitemOpen
  \bibfield  {author} {\bibinfo {author} {\bibfnamefont {S.}~\bibnamefont
  {Van~Gisbergen}}, \bibinfo {author} {\bibfnamefont {J.}~\bibnamefont
  {Snijders}}, \ and\ \bibinfo {author} {\bibfnamefont {E.}~\bibnamefont
  {Baerends}},\ }\href@noop {} {\bibfield  {journal} {\bibinfo  {journal}
  {Chem. Phys. Lett.}\ }\textbf {\bibinfo {volume} {259}},\ \bibinfo {pages}
  {599} (\bibinfo {year} {1996})}\BibitemShut {NoStop}%
\bibitem [{\citenamefont {Park}\ and\ \citenamefont
  {Nordlander}(2009)}]{park2009nature}%
  \BibitemOpen
  \bibfield  {author} {\bibinfo {author} {\bibfnamefont {T.-H.}\ \bibnamefont
  {Park}}\ and\ \bibinfo {author} {\bibfnamefont {P.}~\bibnamefont
  {Nordlander}},\ }\href
  {https://www.sciencedirect.com/science/article/pii/S0009261409003169}
  {\bibfield  {journal} {\bibinfo  {journal} {Chem. Phys. Lett.}\ }\textbf
  {\bibinfo {volume} {472}},\ \bibinfo {pages} {228} (\bibinfo {year}
  {2009})}\BibitemShut {NoStop}%
\bibitem [{\citenamefont {Bardhan}\ \emph {et~al.}(2010)\citenamefont
  {Bardhan}, \citenamefont {Mukherjee}, \citenamefont {Mirin}, \citenamefont
  {Levit}, \citenamefont {Nordlander},\ and\ \citenamefont
  {Halas}}]{bardhan2010nanosphere}%
  \BibitemOpen
  \bibfield  {author} {\bibinfo {author} {\bibfnamefont {R.}~\bibnamefont
  {Bardhan}}, \bibinfo {author} {\bibfnamefont {S.}~\bibnamefont {Mukherjee}},
  \bibinfo {author} {\bibfnamefont {N.~A.}\ \bibnamefont {Mirin}}, \bibinfo
  {author} {\bibfnamefont {S.~D.}\ \bibnamefont {Levit}}, \bibinfo {author}
  {\bibfnamefont {P.}~\bibnamefont {Nordlander}}, \ and\ \bibinfo {author}
  {\bibfnamefont {N.~J.}\ \bibnamefont {Halas}},\ }\href
  {https://doi.org/10.1021/jp9095387} {\bibfield  {journal} {\bibinfo
  {journal} {J. Phys. Chem. C}\ }\textbf {\bibinfo {volume} {114}},\ \bibinfo
  {pages} {7378} (\bibinfo {year} {2010})},\ \Eprint
  {http://arxiv.org/abs/https://doi.org/10.1021/jp9095387}
  {https://doi.org/10.1021/jp9095387} \BibitemShut {NoStop}%
\bibitem [{\citenamefont {Halas}\ \emph {et~al.}(2011)\citenamefont {Halas},
  \citenamefont {Lal}, \citenamefont {Chang}, \citenamefont {Link},\ and\
  \citenamefont {Nordlander}}]{halas2011plasmons}%
  \BibitemOpen
  \bibfield  {author} {\bibinfo {author} {\bibfnamefont {N.~J.}\ \bibnamefont
  {Halas}}, \bibinfo {author} {\bibfnamefont {S.}~\bibnamefont {Lal}}, \bibinfo
  {author} {\bibfnamefont {W.-S.}\ \bibnamefont {Chang}}, \bibinfo {author}
  {\bibfnamefont {S.}~\bibnamefont {Link}}, \ and\ \bibinfo {author}
  {\bibfnamefont {P.}~\bibnamefont {Nordlander}},\ }\href {\doibase
  10.1021/cr200061k} {\bibfield  {journal} {\bibinfo  {journal} {Chem. Rev.}\
  }\textbf {\bibinfo {volume} {111}},\ \bibinfo {pages} {3913} (\bibinfo {year}
  {2011})}\BibitemShut {NoStop}%
\bibitem [{\citenamefont {Prodan}\ \emph {et~al.}(2003)\citenamefont {Prodan},
  \citenamefont {Radloff}, \citenamefont {Halas},\ and\ \citenamefont
  {Nordlander}}]{prodan2003hybridization}%
  \BibitemOpen
  \bibfield  {author} {\bibinfo {author} {\bibfnamefont {E.}~\bibnamefont
  {Prodan}}, \bibinfo {author} {\bibfnamefont {C.}~\bibnamefont {Radloff}},
  \bibinfo {author} {\bibfnamefont {N.~J.}\ \bibnamefont {Halas}}, \ and\
  \bibinfo {author} {\bibfnamefont {P.}~\bibnamefont {Nordlander}},\ }\href
  {\doibase 10.1126/science.1089171} {\bibfield  {journal} {\bibinfo  {journal}
  {Science}\ }\textbf {\bibinfo {volume} {302}},\ \bibinfo {pages} {419}
  (\bibinfo {year} {2003})}\BibitemShut {NoStop}%
\bibitem [{\citenamefont {Nordlander}\ \emph {et~al.}(2004)\citenamefont
  {Nordlander}, \citenamefont {Oubre}, \citenamefont {Prodan}, \citenamefont
  {Li},\ and\ \citenamefont {Stockman}}]{nordlander2004plasmon}%
  \BibitemOpen
  \bibfield  {author} {\bibinfo {author} {\bibfnamefont {P.}~\bibnamefont
  {Nordlander}}, \bibinfo {author} {\bibfnamefont {C.}~\bibnamefont {Oubre}},
  \bibinfo {author} {\bibfnamefont {E.}~\bibnamefont {Prodan}}, \bibinfo
  {author} {\bibfnamefont {K.}~\bibnamefont {Li}}, \ and\ \bibinfo {author}
  {\bibfnamefont {M.~I.}\ \bibnamefont {Stockman}},\ }\href {\doibase
  10.1021/nl049681c} {\bibfield  {journal} {\bibinfo  {journal} {Nano Lett.}\
  }\textbf {\bibinfo {volume} {4}},\ \bibinfo {pages} {899} (\bibinfo {year}
  {2004})}\BibitemShut {NoStop}%
\bibitem [{\citenamefont {Kulkarni}, \citenamefont {Prodan},\ and\
  \citenamefont {Nordlander}(2013)}]{nordlander2013quantum}%
  \BibitemOpen
  \bibfield  {author} {\bibinfo {author} {\bibfnamefont {V.}~\bibnamefont
  {Kulkarni}}, \bibinfo {author} {\bibfnamefont {E.}~\bibnamefont {Prodan}}, \
  and\ \bibinfo {author} {\bibfnamefont {P.}~\bibnamefont {Nordlander}},\
  }\href {\doibase 10.1021/nl402662e} {\bibfield  {journal} {\bibinfo
  {journal} {Nano Lett.}\ }\textbf {\bibinfo {volume} {13}},\ \bibinfo {pages}
  {5873} (\bibinfo {year} {2013})}\BibitemShut {NoStop}%
\bibitem [{\citenamefont {Marinica}, \citenamefont {Aizpurua},\ and\
  \citenamefont {Borisov}(2016)}]{marinica2016quantum}%
  \BibitemOpen
  \bibfield  {author} {\bibinfo {author} {\bibfnamefont {D.-C.}\ \bibnamefont
  {Marinica}}, \bibinfo {author} {\bibfnamefont {J.}~\bibnamefont {Aizpurua}},
  \ and\ \bibinfo {author} {\bibfnamefont {A.~G.}\ \bibnamefont {Borisov}},\
  }\href {\doibase 10.1364/OE.24.023941} {\bibfield  {journal} {\bibinfo
  {journal} {Opt. Express}\ }\textbf {\bibinfo {volume} {24}},\ \bibinfo
  {pages} {23941} (\bibinfo {year} {2016})}\BibitemShut {NoStop}%
\bibitem [{\citenamefont {Thole}(1981)}]{thole1981molecular}%
  \BibitemOpen
  \bibfield  {author} {\bibinfo {author} {\bibfnamefont {B.~T.}\ \bibnamefont
  {Thole}},\ }\href@noop {} {\bibfield  {journal} {\bibinfo  {journal} {Chem.
  Phys.}\ }\textbf {\bibinfo {volume} {59}},\ \bibinfo {pages} {341} (\bibinfo
  {year} {1981})}\BibitemShut {NoStop}%
\end{thebibliography}%

\end{document}